\documentclass[usenatbib]{mn2e}

\usepackage{amsmath}
\usepackage{natbib}
\usepackage{epsfig}
\usepackage{txfonts}
\usepackage{subcaption}
\usepackage{color}
\usepackage{booktabs}
\usepackage{multirow}
\usepackage{epstopdf}

\newcommand{\keV}{{\rm keV}}

\newcommand{\Kel}{{\rm K}}

\newcommand{\Mpc}{{\rm Mpc}}
\newcommand{\GHz}{{\rm GHz}}

\newcommand{\expf}[1]{{{\rm e}^{#1}}}

\newcommand{\beq}{\begin{equation}}   %

\newcommand{\eeq}{\end{equation}}   %

\newcommand{\beqa}{\begin{eqnarray}}   %

\newcommand{\eeqa}{\end{eqnarray}}   %

\newcommand{\beal}{\begin{align}}
\newcommand{\enal}{\end{align}}

\newcommand{\bspl}{\begin{split}}

\newcommand{\espl}{\end{split}}

\newcommand{\bsub}{\begin{subequations}}

\newcommand{\esub}{\end{subequations}}

\newcommand{\bmulti}{\begin{multline}}   %

\newcommand{\beqm}{\begin{mathletters}}   %

\newcommand{\eeqm}{\end{mathletters}}   %

\newcommand{\me}{m_{\rm e}}

\newcommand{\Te}{T_{\rm e}}

\newcommand{\The}{\theta_{\rm e}}

\newcommand{\pot}[2]{#1 \times 10^{#2}}

\usepackage{hyperref}

\title[{\it PIXIE} Spectral Distortion Forecast]
{Prospects for Measuring Cosmic Microwave Background Spectral Distortions in the Presence of Foregrounds}

\author[Abitbol et al.]{Maximilian~H.~Abitbol$^1$\thanks{E-mail:mha2125@columbia.edu}, Jens~Chluba$^2$\thanks{E-mail:jens.chluba@manchester.ac.uk}, J.~Colin~Hill$^3$, and Bradley~R.~Johnson$^1$
\\
$^{1}$ Department of Physics, Columbia University, New York, NY, 10027, USA
\\
$^{2}$ Jodrell Bank Centre for Astrophysics, University of Manchester, Oxford Road, Manchester M13 9PL, UK
\\
$^{3}$ Department of Astronomy, Columbia University, Pupin Hall, New York, New York 10027, USA
}

\begin{document}

\date{\vspace{-3mm} {Accepted 2017 June 20. Received 2017 May 03. Published 2017 June 30. }}
\maketitle

\begin{abstract}
Measurements of cosmic microwave background spectral distortions have profound implications for our understanding of physical processes taking place over a vast window in cosmological history. Foreground contamination is unavoidable in such measurements and detailed signal-foreground separation will be necessary to extract cosmological science. In this paper, we present MCMC-based spectral distortion detection forecasts in the presence of Galactic and extragalactic foregrounds for a range of possible experimental configurations, focusing on the {\it Primordial Inflation Explorer} ({\it PIXIE}) as a fiducial concept. We consider modifications to the baseline {\it PIXIE} mission (operating $\simeq 12$ months in distortion mode), searching for optimal configurations using a Fisher approach. Using only spectral information, we forecast an extended {\it PIXIE} mission to detect the expected average non-relativistic and relativistic thermal Sunyaev-Zeldovich distortions at high significance (194$\sigma$ and 11$\sigma$, respectively), even in the presence of foregrounds. The $\Lambda$CDM Silk damping $\mu$-type distortion is not detected without additional modifications of the instrument or external data. Galactic synchrotron radiation is the most problematic source of contamination in this respect, an issue that could be mitigated by combining {\it PIXIE} data with future ground-based observations at low frequencies ($\nu\lesssim 15-30\,\GHz$). Assuming moderate external information on the synchrotron spectrum, we project an upper limit of $|\mu| < 3.6\times 10^{-7}$ (95\% c.l.), slightly more than one order of magnitude above the fiducial $\Lambda$CDM signal from the damping of small-scale primordial fluctuations, but a factor of $\simeq 250$ improvement over the current upper limit from {\it COBE/FIRAS}. This limit could be further reduced to $|\mu| < 9.4\times 10^{-8}$ (95\% c.l.) with more optimistic assumptions about extra low-frequency information and would rule out many alternative inflation models as well as provide new constraints on decaying particle scenarios.
\end{abstract}


\vspace{-0mm}
\section{Introduction}
\label{sec:Intro}
Spectral distortions of the cosmic microwave background (CMB) are one of the next frontiers in CMB science \citep{Chluba2011therm, Sunyaev2013, Tashiro2014, Silk2014, Chluba2014Moriond, deZotti2015}. The intensity spectrum of the CMB was precisely measured by {\it COBE/FIRAS} over two decades ago~\citep{Mather1990,Fixsen1996} and is consistent with a blackbody at temperature $T_0 = 2.72548 \pm 0.00057$ K from $\nu\simeq3\,\GHz$ to 3,000 GHz~\citep{Fixsen2009}. This agrees with a variety of other CMB observations including COBRA, TRIS, and ARCADE~\citep{gush_halpern_wishnow_rocket,tris_spectraldistortions,Kogut2006ARCADE,arcade2}. 
These impressive measurements already place very tight constraints on the thermal history of the Universe \citep[e.g.,][]{Sunyaev1970SPEC, Zeldovich1972, Illarionov1974, Danese1977}, limiting early energy release to $\Delta\rho_\gamma/\rho_\gamma\lesssim \pot{6}{-5}$ (95\% c.l.) relative to the CMB energy density \citep{Fixsen1996, Fixsen2002}.

Most of our detailed current cosmological picture stems from measurements of the CMB temperature and polarization anisotropies, which have been well characterized by WMAP~\citep{wmap9params}, {\it Planck}~\citep{Planck2015params}, and many sub-orbital experiments \citep[e.g.,][]{Netterfield2002, VSA2003, CBI03, acbar, spt, act}. More precise measurements of the polarization anisotropies are being targeted by an array of ongoing experiments including BICEP2/3 and the KECK array, ACTPol, SPTPol, POLARBEAR, CLASS, SPIDER, and the Simons Array~\citep{bicep2keck_degree,actpol_polresults,sptpol_results,class2016,spider2011}, all with the goal to further refine our understanding of the Universe and its constituents. Measurements of CMB spectral distortions could complement these efforts and provide access to qualitatively new information that cannot be probed via the angular anisotropy \citep[see][for an overview of the standard $\Lambda$CDM distortions]{Chluba2016}. 

\begin{figure*}
\includegraphics[width=0.89\textwidth]{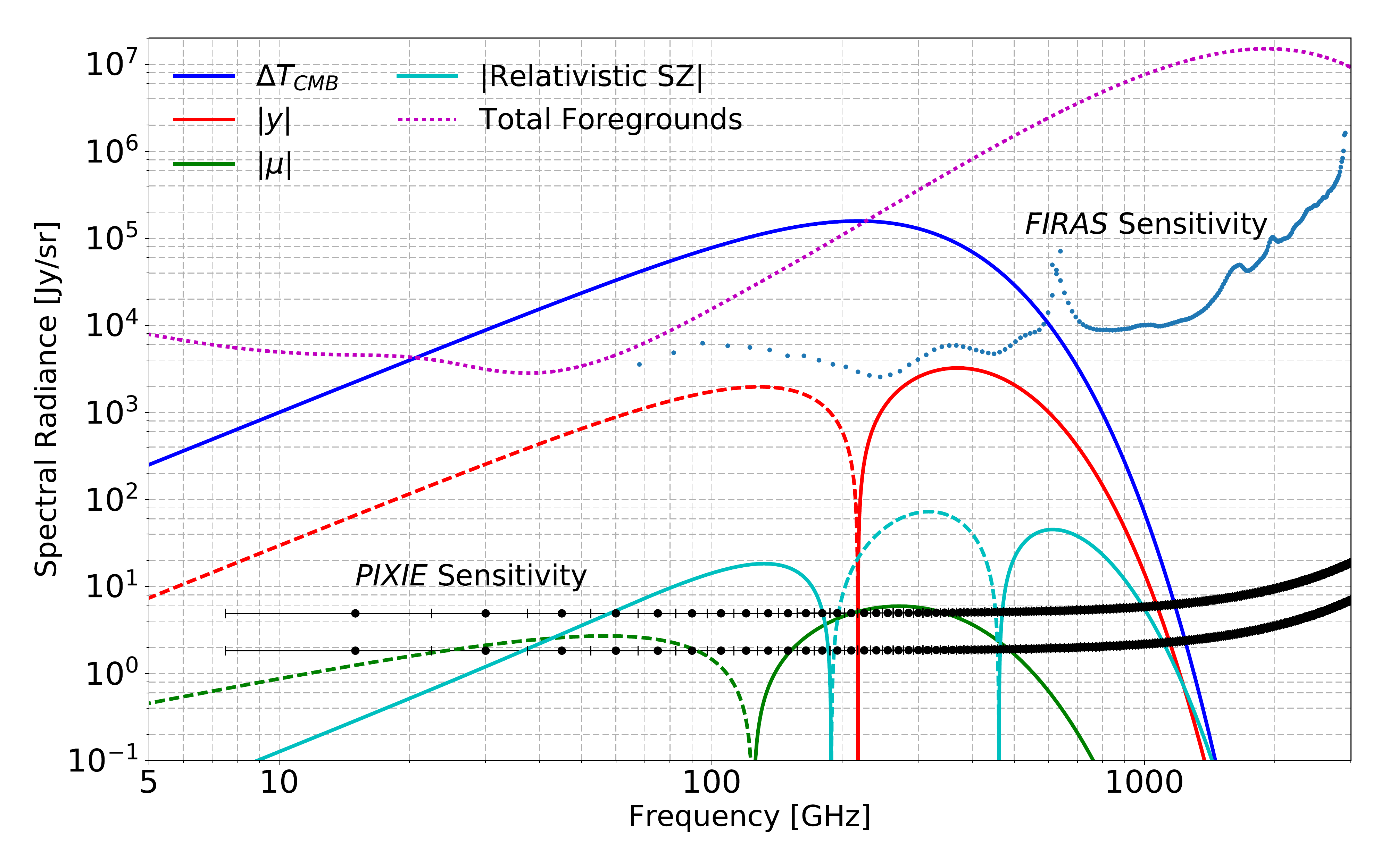}
  \caption{Spectral distortion signals compared to the {\it PIXIE} sensitivity and foregrounds. The signals include the CMB blackbody (blue) as well as the $\Lambda$CDM-predicted Compton-$y$ (red), relativistic SZ (cyan), and $\mu$ (green) spectral distortions. The dashed and solid curves indicate negative and positive values, respectively. The total foreground model, which dominates over all non-blackbody signals, is shown in dotted magenta. The black points represent the {\it PIXIE} sensitivity for the nominal and extended mission, assuming $f_{\rm sky}=0.7$ and 12 or 86.4 months of integration time, respectively. The horizontal error bars on the noise curve points represent the width of the 15~GHz {\it PIXIE} frequency bins. For comparison, the {\it COBE/FIRAS} raw detector sensitivity is illustrated by the blue dots.}
  \label{fig:sdsignals}
\end{figure*}

Spectral distortions are caused by processes (e.g., energy or photon injection) that affect the thermal equilibrium between matter and radiation \citep{Zeldovich1969, Sunyaev1970mu, Illarionov1974, Burigana1991, Hu1993, Chluba2011therm, Chluba2015GreensII}. One of the standard distortions, known as the Compton $y$-distortion, is created in the regime of inefficient energy transfer between electrons and photons, relevant at redshift $z\lesssim \pot{5}{4}$. Processes creating this type of distortion include the inverse-Compton scattering of CMB photons off hot electrons during the epoch of {\it reionization and structure formation} \citep{Sunyaev1972b, Hu1994pert, Refregier2000, Hill2015}, also known in connection with the thermal Sunyaev-Zeldovich (tSZ) effect~\citep{Zeldovich1969}, but can also be related to non-standard physics, e.g., the presence of {\it long-lived decaying particles} \citep{Sarkar1984, Ellis1992, Hu1993b, Chluba2013fore}. 

Chemical potential or $\mu$-type distortions \citep{Sunyaev1970mu}, on the other hand, are generated by energy release at earlier stages ($z\gtrsim \pot{5}{4}$), when interactions are still extremely efficient and able to establish kinetic equilibrium between electrons and photons under repeated Compton scattering and photon emission processes (i.e., double Compton and Bremsstrahlung). The latter are particularly important at $z\gtrsim \pot{2}{6}$, leading to a strong suppression of the distortion amplitude \citep[e.g.,][]{Hu1993}. Expected sources of $\mu$-distortions include the {\it Silk damping of small-scale acoustic modes} in the early Universe~\citep{Sunyaev1970diss, Daly1991, Hu1994, Chluba2012} and the extraction of energy from the photon bath due to the adiabatic cooling of ordinary matter~\citep{Chluba2005, Chluba2011therm}. 

The {\it Primordial Inflation Explorer} ({\it PIXIE}) is a proposed satellite mission designed to constrain the primordial $B$-mode polarization power spectrum and target spectral distortions of the CMB~\citep{Kogut2011,pixie2016}. The instrument is a polarizing Michelson interferometer spanning 15 to 6,000 GHz with a mirror stroke length corresponding to $\simeq 15\,\GHz$ channels. In order to detect the small spectral distortion signals, Galactic and extragalactic foreground emission has to be precisely modeled, characterized, and marginalized over in the cosmological analysis. In this paper we forecast the capabilities of {\it PIXIE}-like experimental concepts for detecting spectral distortions in the presence of known foregrounds, extending simpler forecasts presented earlier \citep[e.g., by][]{Kogut2011, Chluba2013PCA, Hill2015}.\footnote{Forecasts for the polarization sensitivity of {\it PIXIE} were described most recently in~\cite{calabrese_pixieforecast} and will not be addressed here.}  

Our analysis solely considers the spectral energy distribution (SED) of the sky monopole, relying on the spectral behavior of different components in order to separate them.  In contrast, we note that the {\it COBE/FIRAS} analysis relied on spatial information in order to separate the extragalactic monopole from Galactic foregrounds [and the CMB dipole]~\citep{Fixsen1996}.  An optimal analysis would combine spectral and spatial information, with the latter primarily helping to isolate Galactic foregrounds. We shall leave a more rigorous assessment to future work and for now focus on the available spectral information.

We apply a Fisher matrix approach to the fiducial {\it PIXIE} instrument configuration, spectral distortion signals, and standard foreground models to estimate uncertainties on the CMB signal parameters. We consider a range of foreground models and vary the {\it PIXIE} mission configuration to search for an optimal instrument setup based on the assumed sky signals. We compare part of the results to full Monte Carlo Markov Chain (MCMC) analyses, which do not rely on the assumption of Gaussian posteriors, finding good agreement. The considered signals and total foreground emission are illustrated in Fig.~\ref{fig:sdsignals} and will be discussed in detail below.

The paper is organized as follows. We describe the {\it PIXIE} mission, fiducial CMB spectral distortions, and foreground models in Sections~\ref{sec:pixie_config}, \ref{sec:spectral_distortions}, and \ref{sec:foregrounds}, respectively. We summarize the forecasting calculations in Section~\ref{sec:calculation}. The CMB-only forecast is presented in Section~\ref{sec:noiselimited}. Forecasts with foregrounds are discussed in Section~\ref{sec:foregroundforecast}. We search for an optimal mission configuration in Section~\ref{sec:optimal} and conclude in Section~\ref{sec:conclusions}. 

\vspace{-3mm}
\section{{\it PIXIE} Mission Configuration}
\label{sec:pixie_config}
We use the nominal {\it PIXIE} mission configuration as described in~\cite{Kogut2011PIXIE}. (A slightly updated concept was recently proposed\footnote{Al Kogut, priv. comm.} but the modifications do not significantly change the forecast.) The center of the lowest frequency bin is 15~GHz, with a corresponding 15~GHz bin width. The highest frequency bin in the nominal design is $\simeq 6$~THz; however, due to the complexity of dust emission at such high frequencies, we use 3~THz as the highest bin edge, yielding a total of 200 channels for distortion science. This choice does not affect the forecasted uncertainties because the high-frequency foregrounds are not the limiting factor. In addition, the spectral distortion signals cut off well below 3~THz (see Fig.~\ref{fig:sdsignals}). 

Assuming 12 months of spectral distortion mode integration time ({\it PIXIE} will also spend time in polarization observation mode), the noise per $1^{\circ}\times 1^{\circ}$ pixel is $\simeq 747$ Jy at low frequencies, which we convert to Jy/sr and assume $70\%$ of the sky is used in the analysis. The projected noise increases at frequencies above 1~THz due to a low-pass filter in the instrument. For most of the forecasting below, we scale the sensitivity to an extended mission with 86.4 months of integration time (representing a 9 year mission with 80\% of the observation time spent in distortion mode), with the noise scaling down as the square root of the mission duration. For a Fourier Transform Spectrometer (FTS) such as {\it PIXIE}, the lowest frequency bin is set by the size of the instrument. The mirror stroke length determines the frequency resolution. Additionally, increasing the bin width (i.e., reducing the mirror stroke) by a multiplicative factor decreases the noise by the same factor. This is due to the fact that the noise scales with the square root of the bandwidth and the square root of the integration time, both of which increase when increasing the bin size. We discuss the trade-off between frequency resolution and sensitivity in Section~\ref{sec:optimal}.

\vspace{-3mm}
\section{CMB Spectral Distortion Modeling}
\label{sec:spectral_distortions}
At the level of {\it PIXIE}'s expected sensitivity, the average CMB spectral distortion signal can be efficiently described by only a few parameters. We model the sky-averaged spectral radiance relative to the assumed CMB blackbody, $\Delta I_{\nu}$, as:
\begin{equation}
\label{eq.Inu}
\Delta I_{\nu} = \Delta B_{\nu} + \Delta I_{\nu}^y + \Delta I_{\nu}^{\rm rel-tSZ} + \Delta I_{\nu}^{\mu} + \Delta I_{\nu}^{\rm fg} \,.
\end{equation}
Here, $\Delta B_{\nu}=B_\nu\left(T_{\rm CMB}\right)-B_\nu(T_0)$ represents the deviation of the true CMB blackbody spectrum, $B_\nu(T_{\rm CMB})$, at a temperature $T_{\rm CMB}=T_0(1+\Delta_T)$, from that of a blackbody with temperature $T_0=2.726\,\Kel$; $\Delta I_{\nu}^y$ is the $y$-type distortion; $\Delta I_{\nu}^{\rm rel-tSZ}$ is the relativistic temperature correction to the tSZ distortion; $\Delta I_{\nu}^{\mu}$ is the $\mu$-type distortion; and $\Delta I_{\nu}^{\rm fg}$ represents the sum of all foreground contributions. We describe our fiducial models for these signal components below. The results are shown in Figure~\ref{fig:sdsignals} in comparison to the {\it PIXIE} (nominal/extended mission) sensitivity and the total foreground level (described in Sect.~\ref{sec:foregrounds}). 

\vspace{-0mm}
\paragraph*{Blackbody Component.}
The average CMB blackbody temperature must be determined in the analysis, as it is not currently known at the necessary precision \citep[e.g., see][]{Chluba2013PCA}. We work to first order in $\Delta_T=(T_{\rm CMB}-T_0)/T_0$, describing the temperature shift spectrum as
\begin{equation}
\label{eq.Bnu}
\Delta B_{\nu} \approx I_o \frac{x^4 {\rm e}^x}{({\rm e}^x-1)^2} \Delta_T \,,
\end{equation}
with $I_o=(2h/c^2)\,(kT_0/h)^3\approx 270\,{\rm MJy/sr}$ and $x=h\nu/k T_0$. For illustration, we assume a fiducial value $\Delta_T = 1.2 \times 10^{-4}$, consistent with current constraints \citep{Fixsen2009}. The analysis is not affected significantly by this choice. 

While simple estimates indicate that {\it PIXIE} is expected to measure $T_{\rm CMB}$ to the $\simeq {\rm nK}$ level \citep{Chluba2013PCA}, an improvement over {\it COBE/FIRAS} does not immediately provide new cosmological information simply because there is no cosmological prediction for the average photon temperature. By comparing the local ($\leftrightarrow$ current) value of $T_{\rm CMB}$ with measurements at earlier times, e.g., at recombination \citep{Planck2015params} or during BBN \citep{Steigman2009}, constraints on entropy production can be deduced \citep{Steigman2009, Jeong2014}; however, these are not limited by the current $\simeq {\rm mK}$ uncertainty of $T_{\rm CMB}$.

\vspace{-0mm}
\paragraph*{Cumulative Thermal SZ ($y$) Distortion.}
We adopt the model for the sky-averaged thermal SZ signal from~\cite{Hill2015}, including both the standard non-relativistic (Compton-$y$) and relativistic contributions.\footnote{{\it PIXIE} may also have sufficient sensitivity to constrain the sky-averaged {\it non-thermal} SZ signal, but we do not investigate this possibility here.} The Compton-$y$ signal (tSZ) includes contributions from the intracluster medium (ICM) of galaxy groups and clusters (which dominate the overall signal), the intergalactic medium, and reionization, yielding a total value of $y = 1.77 \times 10^{-6}$~\citep{Hill2015}. This is a conservative estimate as with increased AGN feedback larger values for $y$ could be feasible \citep{deZotti2015}.
Note that the actual monopole $y$ value measured by {\it PIXIE} or other experiments will also contain a primordial contribution in general, but this is expected to be 2--3 orders of magnitude smaller than the structure formation contributions~\citep{Chluba2012}. We furthermore assume that the average $y$-distortion caused by the CMB temperature dipole, $y_{\rm sup}=\pot{(2.525\pm0.012)}{-7}$ \citep{Chluba2004, Chluba2016}, is subtracted.
The non-relativistic tSZ signal takes the standard Compton-$y$ form~\citep{Zeldovich1969}:
\begin{equation}
\label{eq.Inu_y}
\Delta I_{\nu}^y = I_o \frac{x^4 {\rm e}^{x}}{\left({\rm e}^{x} - 1\right)^2} \left[ x \coth\left( \frac{x}{2} \right) - 4 \right] \,y \,,
\end{equation}
with cross-over frequency $\nu \simeq 218\,\GHz$.

We model the sky-averaged relativistic corrections to the tSZ signal, $\Delta I_{\nu}^{\rm rel-tSZ}$, using the moment-based approach described in~\cite{Hill2015}, whose calculation used the results of~\cite{Nozawa2006} up to fourth order in the electron temperature. For the MCMC calculations below, we generate the signal using moments of the optical-depth-weighted ICM electron temperature distribution of {\tt SZpack} \citep{Chluba2012moments}, with parameter values identical to those in~\cite{Hill2015}, to which we refer the reader for more details.
However, at {\it PIXIE}'s sensitivity, the SZ signal can be represented most efficiently using moments of the $y$-weighted ICM electron temperature distribution. In particular, using only the first two $y$-weighted moments is sufficient to reproduce the relativistic correction signal for our purpose. This is explained in more detail by \citet{Battaglia2016} and greatly simplifies comparisons to the results of cosmological hydrodynamics simulations. 

While we emphasize that the fiducial signal is generated using the more accurate optical-depth-weighted approach (in the MCMC case), the Fisher forecasts and MCMC fits below use the $y$-weighted moment approach in the analysis. The two approaches are equivalent in the limit of many temperature moments, but to reduce the number of parameters, the $y$-weighted approach provides an efficient re-summation of the signal templates. We denote the first moment of the $y$-weighted ICM electron temperature distribution as $kT_{\rm eSZ}$. The fiducial value is $kT_{\rm eSZ} = 1.245\,\keV$, which is recovered in noiseless estimates of the full signal (including all higher temperature moments) for {\it PIXIE} channel settings. 

One can think of all-sky SZ observations as the ultimate stacking method for SZ halos. In foreground-free forecasts, the second moment of the underlying relativistic electron temperature distribution, $\omega_2^{\rm eSZ}$, is also detectable with an extended {\it PIXIE} mission (see Table~\ref{tab:cmbonly}). In this case, the recovered noiseless relativistic correction parameters are $kT_{\rm eSZ} = 1.282\,\keV$ and $\omega_2^{\rm eSZ}=1.152$ (again, for default {\it PIXIE} channel settings).
The spectral templates for the relativistic tSZ signal can be expressed as
\begin{align}
\label{eq.Inu_rel}
\Delta I_{\nu}^{\rm rel-SZ} &= I_o \frac{x^4 {\rm e}^{x}}{\left({\rm e}^{x} - 1\right)^2} \left\{ Y_1(x) \,\The +Y_2(x) \,\The^2+Y_3(x) \,\The^3 
\right.
\nonumber\\
&\qquad+
\left. 
\left[Y_2(x) \,\The^2+3Y_3(x) \,\The^3\right]\,\omega_2^{\rm eSZ}
\right\}\,y
\end{align}
to sufficient precision for our analysis. Here, $\The=kT_{\rm eSZ}/\me c^2$ and $Y_i(x)$ are the usual functions obtained by asymptotic expansions of the relativistic SZ signal \citep{Sazonov1998, Challinor1998, Itoh98}. By characterizing the relativistic tSZ contribution one can learn about feedback processes during structure formation \citep{Hill2015, Battaglia2016}.

\begin{figure*}
	\includegraphics[width=0.90\textwidth]{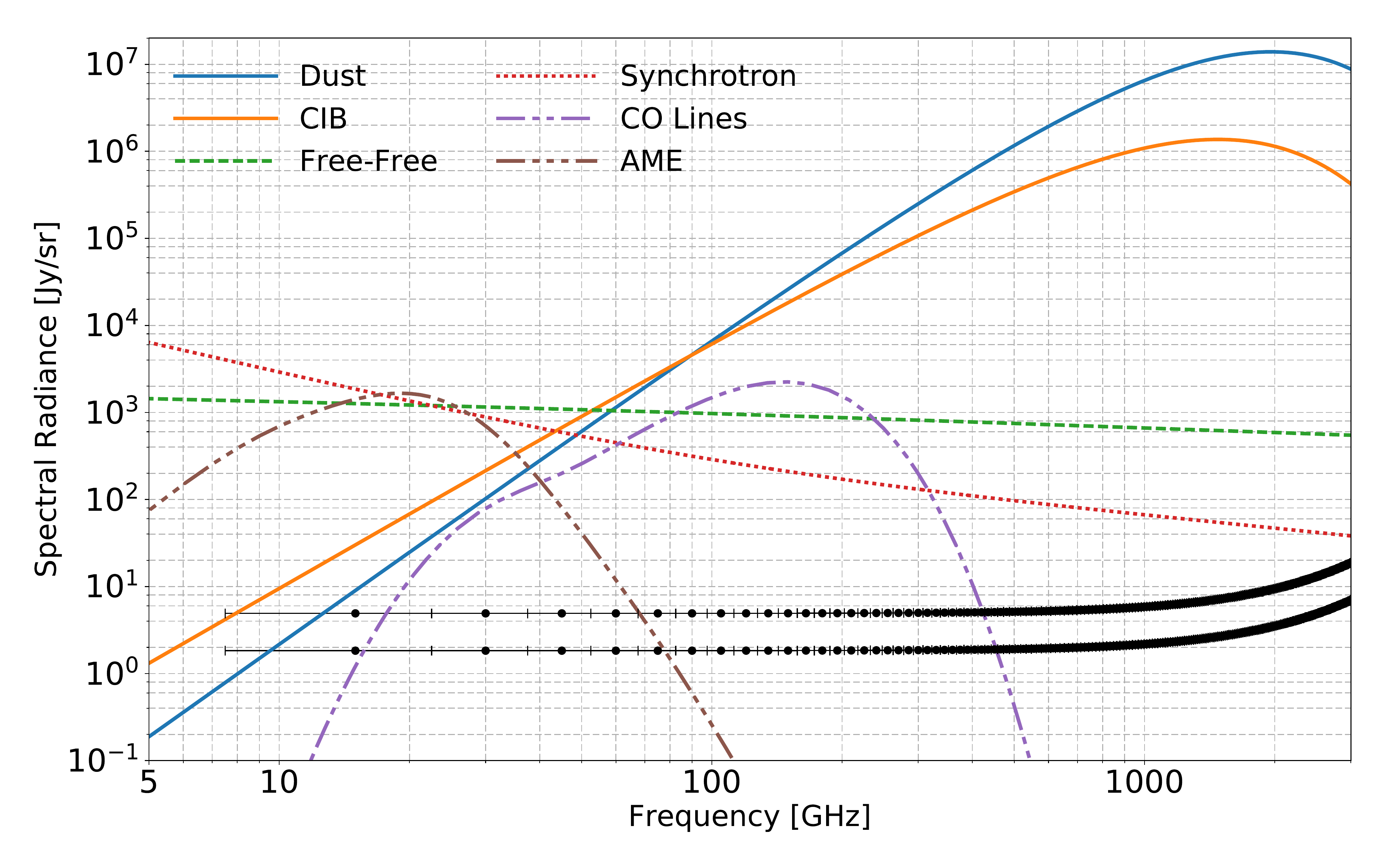}
  \caption{Foreground spectral radiance relative to the CMB blackbody, $\Delta I^{\rm fg}_{\nu}$, for the components in our model (as labeled in the figure). The black points with horizontal error bars represent the {\it PIXIE} sensitivity and bin width for the nominal and extended mission duration, as in Figure~\ref{fig:sdsignals}. Galactic and extragalactic dust emission dominates at high frequencies, but multiple components are important at low frequencies ($\nu \lesssim 100$ GHz).}
  \label{fig:fgsignals}
\end{figure*}

\vspace{3mm}
\paragraph*{Primordial $\mu$ Distortion.} 
Chemical potential $\mu$-type distortions \citep{Sunyaev1970mu} can be generated by many forms of energy release at redshifts $5\times 10^4 \lesssim z \lesssim 2 \times 10^6$, including decaying or annihilating particles \citep[e.g.,][]{Sarkar1984,Hu1993b,McDonald2001, Chluba2013fore}, the damping of small-scale density fluctuations \citep[e.g.,][]{Sunyaev1970diss,Daly1991,Barrow1991,Chluba2012}, and injection from cosmic strings \citep{Ostriker1987, Tashiro2012,Tashiro2012b} or primordial black holes \citep{Carr2010, Yacine2016}. A negative $\mu$ distortion is also generated by the Compton-cooling of CMB photons off the adiabatically evolving electrons \citep{Chluba2005, Chluba2011therm}. 

Here, we assume only the ``vanilla'' sources exist in our Universe, in particular the $\mu$ signals from acoustic damping and adiabatic cooling~\citep{Chluba2012, Chluba2016}. The latter signal is expected to be roughly one order of magnitude smaller than the former. We adopt a fiducial value of $\mu = 2 \times 10^{-8}$, consistent with current constraints on the primordial power spectrum \citep{Chluba2012, Chluba2013PCA, Cabass2016, Chluba2016}. The spectral dependence of the $\mu$-distortion is given by \citep[e.g.,][]{Chluba2012, Chluba2013Green}:
\begin{equation}
\label{eq.Inu_mu}
\Delta I_{\nu}^{\mu} = I_o \frac{x^4 {\rm e}^x}{({\rm e}^x-1)^2} \left[ \frac{1}{\beta} - \frac{1}{x} \right] \,\mu \,.
\end{equation} 
with $\beta\approx 2.1923$. This distortion has a shape that is similar to that of the $y$-type distortion, but with a zero-crossing at $\nu\simeq 125\,\GHz$ rather than $\nu\simeq 218\,\GHz$ (see Fig.~\ref{fig:sdsignals}). By measuring the $\mu$-distortion parameter one can place tight limits on the amplitude of the primordial power spectrum at small scales corresponding to wavenumber $k\simeq 10^3\,\Mpc^{-1}$ \citep[e.g.,][]{Chluba2012, Chluba2012inflaton, Clesse2014}.

\vspace{-0mm}
\paragraph*{Residual Distortions.}
For a given energy release history, in general spectral distortions are generated which are not fully described by the sum of the $\mu$- and $y$-type shapes~\citep{Chluba2011therm,Khatri2012mix,Chluba2013Green}, yielding the  so-called ``residual'' or $r$-type distortion. However, for the concordance $\Lambda$CDM cosmology, the lowest-order $r$-type distortion is expected to be well below {\it PIXIE}'s sensitivity \citep{Chluba2013PCA,Chluba2016}.  Thus, we neglect this contribution in the forecasts carried out below. In an analysis of actual {\it PIXIE} data, it will be interesting to search for and constrain this signal, as it can be sizeable in non-standard scenarios, e.g., those related to energy release from decaying particles \citep{Chluba2013PCA}.

\vspace{-0mm}
\paragraph*{Cosmological Recombination Spectrum.}
The cosmological recombination process occurring at $z\simeq 10^3$ also causes a very small distortion of the CMB spectrum visible at $\nu \simeq 1\,\GHz - 3\,{\rm THz}$ through photon injection \citep{Zeldovich68, Dubrovich1975, Sunyaev2009}. This signal can now be accurately computed \citep{Jose2006, Chluba2006b, Chluba2010, Yacine2013RecSpec, Chluba2016CosmoSpec} and would provide a novel way to constrain cosmological parameters, such as the baryon density and primordial helium abundance \citep{Chluba2008T0, Sunyaev2009}. However, since this signal is about one order of magnitude below the sensitivity of {\it PIXIE} \citep{Vince2015}, we neglect it in our analysis. A detection of the recombination signal could become feasible in the future using ground-based detector arrays operating at low frequencies, $\nu\simeq 2-6\,\GHz$ \citep{Mayuri2015}.

\begin{table*}
\centering
\caption{Foreground model motivated by {\it Planck} data. All SEDs, $\Delta I_{\rm X}$, are in units of Jy/sr. For each component, we also give the value of $\Delta I_{\rm X}(\nu_r)$ at $\nu_r=100\,\GHz$ for reference.}
\begin{tabular}{lccccc}
\hline
\hline
 Foreground 
 & Spectral Radiance [Jy/sr]
 & Free Parameters and Values
 & Additional Information
 \\
\hline
\hline
\multirow{3}{*}{Thermal Dust}
& $x = \frac{h \nu}{k T_{\rm D}}$
& $A_{\rm D} = 1.36 \times 10^6$~Jy/sr
& $\Delta I_{\rm D}(\nu_r) = 6,608$~Jy/sr
\\[1mm]
& $\Delta I_{\rm D}(\nu) = A_{\rm D}\,x^{\beta_{\rm D}}\, \frac{x^3}{\expf{x}-1}$ 
& $\beta_{\rm D} = 1.53$
& 
\\[1mm]
& 
& $T_{\rm D} = 21$~K
& 
\\
\hline
\multirow{3}{*}{CIB}
& $x = \frac{h \nu}{k T_{\rm CIB}}$
& $A_{\rm CIB} = 3.46 \times 10^5$~Jy/sr
& $\Delta I_{\rm CIB}(\nu_r) = 6,117$~Jy/sr
\\[1mm]
& $\Delta I_{\rm CIB}(\nu) = A_{\rm CIB}\,x^{\beta_{\rm CIB}}\, \frac{x^3}{\expf{x}-1}$ 
& $\beta_{\rm CIB} = 0.86$
& 
& \\[1mm]
& 
& $T_{\rm CIB} = 18.8$~K
& 
\\
\hline
\multirow{3}{*}{Synchrotron}
&
& $A_{\rm S} = 288.0$~Jy/sr
& $\Delta I_{\rm S}(\nu_r) = 288$~Jy/sr
\\[1mm]
& $\Delta I_{\rm S}(\nu) = A_{\rm S} \left(\frac{\nu}{\nu_0}\right)^{\alpha_{\rm S}} \Big[1 + \frac{1}{2} \omega_{\rm S} \ln^2 \left(\frac{\nu}{\nu_0}\right) \Big]$
& $\alpha_{\rm S} = -0.82$
& $10\%$ prior assumed on $A_{\rm S}$ and $\alpha_{\rm S}$ \\[1mm]
& 
& $\omega_{\rm S} = 0.2$
& $\nu_0 = 100$~GHz 
\\
\hline
\multirow{2}{*}{Free-Free}
& $\nu_{\rm ff} = \nu_{\rm FF}\,(\Te/10^{3}~\textnormal{K})^{3/2}$
& $A_{\rm FF} = 300$~Jy/sr
& $\Delta I_{\rm FF}(\nu_r) = 972$~Jy/sr
\\[1mm]
& $\Delta I_{\rm FF}(\nu) = A_{\rm FF}\left(1 + \ln\Big[1 + \left(\frac{\nu_{\rm ff}}{\nu}\right)^{\sqrt{3}/\pi}\Big]\right)$
& 
& $\{\Te, \nu_{\rm FF}\} 
= \{7000$~K, 255.33\,\GHz \} 
\\
\hline
\multirow{2}{*}{Integrated CO}
& $\Theta_{\rm CO}(\nu) = \textnormal{CO template}(\nu)$
& $A_{\rm CO} = 1$
& $\Delta I_{\rm CO}(\nu_r) = 1,477$~Jy/sr
\\[1mm]
& $\Delta I_{\rm CO}(\nu) = A_{\rm CO} \Theta_{\rm CO}(\nu)$
& 
& Template in Jy/sr
\\
\hline
\multirow{2}{*}{Spinning Dust}
& $\Theta_{\rm SD}(\nu) = \textnormal{SD template}(\nu)$
& $A_{\rm SD} = 1$
& $\Delta I_{\rm SD}(\nu_r) = 0.25$~Jy/sr
\\[1mm]
& $\Delta I_{\rm SD}(\nu) = A_{\rm SD} \Theta_{\rm SD}(\nu)$
& 
& Template in Jy/sr
\\[1pt]
\hline
\hline
\label{tab:foregroundfunctions}
\end{tabular}
\end{table*}

\vspace{-2mm}
\section{Foreground Modeling}
\label{sec:foregrounds}
We consider six main astrophysical foregrounds which contaminate CMB measurements in the frequency range from 10~GHz to 3~THz: Galactic thermal dust, cosmic infrared background (CIB), synchrotron, free-free, integrated CO, and spinning dust emission (anomalous microwave emission, or AME). We use the {\it Planck} results to estimate each component's SED~\citep{planck2015Xforegrounds}. While {\it Planck} measured intensity fluctuations, {\it PIXIE} will measure the absolute sky intensity. We assume that the SEDs of the fluctuations measured by {\it Planck} can be used to model the monopole SED for each foreground. 

Figure~\ref{fig:fgsignals} shows each foreground SED and Table~\ref{tab:foregroundfunctions} lists the relevant parameters. All SEDs are given in absolute intensity units of spectral radiance [Jy/sr]. The results presented below do not depend strongly on the amplitudes of the foregrounds (within reasonable ranges), but they do depend on the spectral shapes of the foregrounds and the associated number of free parameters (see Sect.~\ref{sec:foreground_model} for discussion). The foreground-to-signal level sets the calibration requirement for the instrument, but we ignore instrumental systematics for the purposes of this work. Our results hold as long as the shapes of the foreground emission do not deviate strongly from the assumed models. 

An important note is that for this analysis we assume these SEDs represent the sky-averaged spectra of the foregrounds. The foreground emission varies across the sky and the average of a specific SED model over a distribution of parameter values is not in general represented by the same SED with a single set of parameter values. For example, averaging many modified blackbody spectra with different temperatures does not correspond exactly to a modified blackbody with a single temperature [similarly, averaging over the CMB temperature anisotropy itself produces a small $y$-distortion~\citep{Chluba2004}]. A moment expansion approach could appropriately handle the effects of different averaging processes~\citep{Chluba2017}. Spatial information about the Galactic foreground emission could also help in separating signals from foregrounds. We limit the scope of this paper to understanding how the shapes of known foreground SEDs impact our ability to measure spectral distortions and leave spatial considerations to future work. In addition, we do not consider effects due to imperfect modeling, as described for example in~\cite{caimapo_foregrounds} for a polarization forecast.

For {\it PIXIE}, we find below that the limiting foregrounds are those which dominate at low frequency, specifically synchrotron and free-free emission, and spinning dust to a lesser extent. Additional measurements of these signals below 100~GHz will be necessary for a detection of the standard $\Lambda$CDM $\mu$-type spectral distortion. {\it PIXIE} will set the most stringent constraints to date on the thermal dust and CIB emission and these limits could only be improved by increasing the effective integration time of the experiment. We consider including information from external datasets in the form of priors on select foreground parameters in Section~\ref{sec:priors}. In total there are 12 free foreground parameters. We describe our model for each foreground SED in the following.

\vspace{-1.5mm}
\paragraph*{Thermal Dust and Cosmic Infrared Background.}
The brightest foregrounds at frequencies above 100~GHz are due to Galactic thermal dust and the cumulative redshifted emission of thermal dust in distant galaxies, called the cosmic infrared background (CIB). The physical characteristics of dust grains, such as the molecular composition, grain size, temperature, and emissivity, vary widely in the Galaxy, but for CMB analyses this is often summed into a modified blackbody spectrum. {\it Planck} finds empirically that a single-temperature modified blackbody describes the observed emission well, and we therefore adopt this model~\citep{planck2015Xforegrounds}. We similarly use a modified blackbody to represent the CIB emission, and use data from {\it Planck} to determine the parameters~\citep{planck2013XXXcib}. Due to the more complex emission and absorption spectra of dust at frequencies near and above 1~THz, the {\it Planck} analysis cautions against the use of the model at such high frequencies. However, for the purpose of this forecast we extend the model to 3~THz. Cutting the forecast off at 1~THz only marginally affects the forecasted {\it PIXIE} performance. 

Each modified blackbody is characterized by 3 parameters: the amplitude, spectral index, and dust temperature, for a total of 6 thermal dust and CIB foreground parameters. This list can be extended using a moment expansion method \citep{Chluba2017}; however, in this case the use of spatial information, possibly from future high-resolution CMB imagers \citep{COrE2011, PRISM2013WPII}, is essential but beyond the scope of this work, so that we limit ourselves to a 6-parameter model.

\vspace{-2mm}
\paragraph*{Synchrotron.} 
The most dominant low-frequency foreground comes from the synchrotron emission of relativistic cosmic ray electrons deflected by Galactic magnetic fields. The shape of this emission is predicted to obey a power law with a spectral index of approximately $\alpha_{\rm sync} \simeq -1$ in intensity units (i.e., approximately $\alpha^T_{\rm sync} \simeq-3$ in brightness temperature). Empirically, {\it Planck} finds a power law with a flattening at low frequencies to best fit the data~\citep{planck2015Xforegrounds}. 
Additional low-frequency SED modeling is discussed in \cite{Oliveira2008} and \cite{Sathyanarayana2017}, with the latter introducing physically-motivated SED approximations.

To avoid the use of a template and allow for a more general SED we use a power law with logarithmic curvature to describe the Galactic synchrotron emission. Such spectral curvature generically arises when averaging over power-law SEDs with different spectral indices \citep[e.g.,][]{Chluba2017}. There are thus 3 free parameters in our model: the amplitude, spectral index, and curvature index. Spectral curvature is usually neglected for single-pixel SED modeling; however, line-of-sight and beam averages cannot be avoided and thus require its inclusion at the level of sensitivity reached by {\it PIXIE} \citep{Chluba2017}.

We estimate the synchroton parameters by fitting this model to the {\it Planck} synchrotron spectrum. Unless stated otherwise, we impose a 10\% prior on the synchrotron amplitude and spectral index throughout the forecasting to represent the use of external datasets such as the Haslam 408~MHz map, {\it WMAP}, {\it Planck}, C-BASS, QUIJOTE, or future observations~\citep{haslam1981,haslam1982,haslamupdate,wmap9results,planck2015Xforegrounds,cbass2015,quijote2015}. We find that in particular future low-frequency ($\lesssim 15-30$~GHz) observations that can constrain the synchrotron SED or limit its contribution using spatial information to better than $1\%$ will be very valuable for tightening the constraints on $\mu$.

\vspace{-2mm}
\paragraph*{Free-free Emission.}
The next brightest foreground at low frequencies, following a relatively shallow spectrum, is the thermal free-free ($\leftrightarrow$ Bremsstrahlung) emission from electron-ion collisions within the Galaxy (for example in HII regions). The shape of the spectrum is derived from~\cite{draineISM}. We neglect the small high-frequency suppression (at $\nu\gtrsim 1\,{\rm THz}$) caused by the presence of CMB photons in the spectral template \citep{Chluba2017}. At the relevant frequencies, the spectrum is very weakly dependent on the electron temperature, and we therefore only allow for one free parameter, corresponding to the overall amplitude in intensity units. We estimate the amplitude by fitting this model to the free-free spectrum from {\it Planck}~\citep{planck2015Xforegrounds}. 

\vspace{-2mm}
\paragraph*{Cumulative CO.}
The cumulative CO emission from distant galaxies adds another foreground that will interfere with CMB spectral distortion measurements. 
From the theoretical point of view, the exact spectral shape is very uncertain and depends on the star-formation history \citep{Righi2008b}. Recent observations with {\it ALMA} place a lower limit on the integrated cosmic CO signal \citep{alma_CII}, which is consistent with these models.

Here, we take the spectra calculated by \cite{Mashian2016} and produce a template with one free amplitude parameter to model the average CO emission. In principle, one could allow the spectral shape of each individual line to vary (with some relative constraints on the amplitudes), but for simplicity we use only one template. We note that cross-correlations with galaxy redshift surveys could provide an independent estimate of the CO SED (and other lines) via intensity mapping \citep{highzISM_CII,switzerCO}, which could be used to improve the modeling.

\vspace{-2mm}
\paragraph*{Spinning Dust Grains.}
Lastly, we consider AME, which is non-negligible at $\nu\simeq 10-60\,\GHz$ and thought to be sourced by spinning dust grains with an electric dipole moment~\citep{draine_ame}. We adopt the model used by {\it Planck}, which generates a template from a theoretically calculated SED~\citep{planck2015Xforegrounds} [see also~\cite{yacine_amereview} and references therein]. We allow one free parameter for the amplitude of the AME template. This is a relatively rigid parameterization, but the AME is not a dominant source of error in the forecast and we find that expanding the model does not significantly change the results. We furthermore anticipate that future ground-based observations will help to improve the modeling of this component, given its potential relevance to ongoing and planned $B$-mode searches \citep{cbass2015,quijote2015}.

\begin{table*}
\centering
\caption{CMB-only MCMC forecasts. This table gives noise-limited constraints for CMB spectral distortion parameters in a no-foreground scenario, derived via MCMC methods. The fiducial parameter values are $\Delta^{\rm f}_T=\pot{1.2}{-4}$, $y^{\rm f}=\pot{1.77}{-6}$, $\mu^{\rm f}=\pot{2.0}{-8}$, and $kT^{\rm f}_{\rm eSZ}=1.245\,\keV$. The table lists the MCMC-recovered values with 1$\sigma$ uncertainties, as well as detection significances in parentheses (fiducial parameter value divided by 1$\sigma$ error). We illustrate the improvement resulting from an extended nine-year {\it PIXIE} mission (86.4 months of integration time in spectral distortion mode). We consider spectral distortion models of increasing complexity to examine potential biases in the parameters. Parameter values that are left blank were not included in the model. The small biases seen in $\mu$ and $\Delta_T$ are due to the fact that the relativistic SZ signal includes contributions from higher-order temperature moments. This bias disappears when including the $y$-weighted temperature dispersion, $\omega_2^{\rm eSZ}$, in the analysis. In this case, one expects to find $kT_{\rm eSZ}\simeq 1.282\,\keV$ and $\omega_2^{\rm eSZ}\simeq1.152$ in noiseless observations (see Sect.~\ref{sec:noiselimited} for more discussion).}
\begin{tabular}{lccccc}
\hline
\hline
 Parameter
 & Baseline 
 & Extended
 & Baseline
 & Extended 
 & Extended
 \\
\hline
\hline
$(\Delta_T-\Delta_T^{\rm f}) \, [10^{-9}]$
& $0.0^{+ 2.3} _{- 2.3}$
& $0.00^{+ 0.85} _{- 0.85}$
& $-0.5^{+ 2.3} _{- 2.3}$
& $-0.53^{+ 0.84} _{- 0.86}$
& $0.00^{+ 0.87} _{- 0.87}$
\\[2pt]
$y\,[10^{-6}]$
& $1.7700 ^{+ 0.0012 } _{- 0.0012 }\,(1475\sigma)$
& $1.77000 ^{+ 0.00044 } _{- 0.00044 }\,(4023\sigma)$
& $1.7692 ^{+ 0.0012 } _{- 0.0012 }\,(1474\sigma)$
& $1.76921 ^{+ 0.00044 } _{- 0.00044 }\,(4021\sigma)$
& $1.76996 ^{+ 0.00050 } _{- 0.00050 }\,(3540\sigma)$
\\[2pt]
$\mu\,[10^{-8}]$
& $2.0 ^{+ 1.3 } _{- 1.3 }\,(1.5\sigma)$
& $2.00 ^{+ 0.50 } _{- 0.50 }\,(4.0\sigma)$
& $2.3^{+ 1.4 } _{- 1.4 }\,(1.6\sigma)$
& $2.30 ^{+ 0.53 } _{- 0.52 }\,(4.3\sigma)$
& $2.00 ^{+ 0.53 } _{- 0.53 }\,(3.8\sigma)$
\\[2pt]
$kT_{\rm eSZ} \,[\keV]$
& $-$
& $-$
& $ 1.244^{+ 0.029 } _{- 0.030 }\,(42\sigma)$
& $1.244 ^{+ 0.011 } _{- 0.011 }\,(113\sigma)$
& $1.281 ^{+ 0.016 } _{- 0.016 }\,(80\sigma)$
\\[2pt]
$\omega_2^{\rm eSZ}$
& $-$
& $-$
& $-$
& $-$
& $1.14 ^{+ 0.32 } _{- 0.33 }\,(3.5\sigma)$
\\[1pt]
\hline
\hline
\label{tab:cmbonly}
\end{tabular}
\end{table*}

\vspace{-0mm}
\paragraph*{Other Components.}
For the purpose of our forecast, we only include the above foregrounds, which are well-known and (relatively) well-characterized.  We neglect several other potential foreground signals, such as additional spectral lines \citep[e.g., CII; see][]{alma_CII, highzISM_CII} or intergalactic dust~\citep{imara_intergalacticdust}. In an effort to capture the dominant effects of the known foregrounds, we also do not include more general models for our foreground signals, with some possible generalizations being discussed in \citet{Chluba2017}. One could also use physical models instead of templates for the CO and AME, or extended dust models with distributions of temperatures and emissivities~\citep[e.g.,][]{kogut_2tempdust, Chluba2017}. This will be studied in the future and also requires taking spectral-spatial information into account.

\vspace{-2mm}
\section{Forecasting Methods}
\label{sec:calculation}
We implement two methods to estimate the capability of {\it PIXIE} (or other CMB spectrometers) to constrain the signals described above. First, we use a Markov Chain Monte Carlo (MCMC) sampler to calculate the parameter posterior distributions. This allows us to determine the most likely parameter values and the parameter uncertainties, even in the case of highly non-Gaussian posteriors, as can be encountered close to the detection threshold. 
Second, we employ a Fisher matrix calculation to determine the parameter uncertainties, assuming Gaussian posteriors. The Fisher method has the benefit of running much more quickly than the MCMC, which allows us to more easily explore the effects of modifying the instrument configuration. In the high-sensitivity limit (i.e., when Gaussianity is an excellent approximation), the two methods converge to identical results. The Fisher information matrix is calculated as
\begin{equation}
F_{ij} = \sum_{a,b} \frac{\partial(\Delta I_{\nu})_a}{\partial p_i} C^{-1}_{ab} \frac{\partial(\Delta I_{\nu})_b}{\partial p_j} \,.
\label{eq:fishermat}
\end{equation}
Here the sum is over frequency bins indexed by $\{a,b\}$, $p_i$ stands for parameter $i$, and $C_{ab}$ is the {\it PIXIE} noise covariance matrix, which we assume to be diagonal. The parameter covariance matrix is then calculated by inverting the Fisher information, $F_{ij}$.

For the MCMC sampling, we use the {\tt emcee} package \citep{Foreman2012}, with wrappers developed previously as part of {\tt SZpack} \citep{Chluba2012moments} and {\tt CosmoTherm} \citep{Chluba2013fore}.
This method allows us to obtain realistic estimates for the detection thresholds when non-Gaussian contributions to the posteriors become noticeable. It also immediately reveals parameter biases introduced by incomplete signal modeling. We typically use $N\simeq 200$ independent walkers and vary the total number of  samples to reach convergence in each case. Unless stated otherwise, flat priors over a wide range around the input values are assumed for each parameter. We impose a lower limit $A_{\rm sync}>0$, as in many of the estimation problems this unphysical region would otherwise be explored due to the large error on $A_{\rm sync}$.
For high-dimensional cases (14 and 16 parameters), we find the convergence of the affine-invariant ensemble sampler in the {\tt emcee} package to become extremely slow, so that in the future alternative samplers should be used.

\vspace{-2mm}
\section{CMB-only distortion sensitivities}
\label{sec:noiselimited}
To estimate the maximal amount of information that {\it PIXIE} could extract given its noise level, we perform several MCMC forecasts omitting foreground contamination. The CMB parameters are $\Delta_T=(T_{\rm CMB}-T_0)/T_0$, $y$, $kT_{\rm eSZ}$, and $\mu$. Considering the cases with only $\Delta_T$, $y$, and $\mu$ (i.e., neglecting the relativistic SZ temperature corrections), the baseline mission ($12$ months spent in distortion mode) yields a significant detection of the $y$-parameter, but only a marginal indication for non-zero $\mu$ (see Table~\ref{tab:cmbonly}). This situation improves for an extended mission ($86.4$ months in distortion mode), suggesting that a $\simeq 4\sigma$ detection of $\mu$ would be possible. In both cases, the constraints are driven by channels at $\nu\lesssim 1\,{\rm THz}$.

When adding the relativistic temperature correction to the SZ signal and modeling the data using $\Delta_T$, $y$, $\mu$, and the $y$-weighted electron temperature $kT_{\rm eSZ}=\left< y \,k\Te\right>/\left<y\right>$, only a small penalty is paid in the constraint on $\mu$: the error increases from $\sigma_{\mu} \simeq \pot{1.3}{-8}$ to $\sigma_{\mu} \simeq \pot{1.4}{-8}$ for the baseline mission, consistent with the results of~\cite{Hill2015}. A very significant measurement of $kT_{\rm eSZ}$ is expected. The central value of $\mu$ is biased high by $\Delta \mu\simeq \pot{0.3}{-8}$, since the relativistic SZ correction  model includes contributions from higher-order moments that are not captured by only adding $kT_{\rm eSZ}$ (see Sect.~\ref{sec:spectral_distortions}). 
When also adding the second moment of the $y$-weighted electron temperature to the analysis ($\omega_2^{\rm eSZ}$), this bias disappears. The main penalty for adding this parameter is paid by $kT_{\rm eSZ}$, for which the detection significance degrades by a factor of $\simeq 1.4$. The second temperature moment is seen at a similar level of significance as $\mu$. The relativistic distortion signal receives extra information from frequencies $\nu\simeq 1\,{\rm THz}-2\,{\rm THz}$, which makes it distinguishable from $\mu$ without impacting its constraint. For baseline settings, we find $\mu=\pot{(2.0\pm 1.4)}{-8}\,(\simeq 1.4\sigma)$, $kT_{\rm eSZ}=(1.279\pm 0.042)\,\keV\,(\simeq 31\sigma)$ and $\omega_2^{\rm eSZ}= 1.12 ^{+ 0.84 } _{- 0.93 }\,(\simeq 1.1\sigma)$, and practically unaltered constraints on $\Delta_T$ and $y$. 
Overall, we conclude that for an extended mission and a {\it foreground-free sky}, the noise level of {\it PIXIE} would be sufficient to detect the standard $\Lambda$CDM $\mu$ distortion at moderate significance, as well as the $y$ and $kT_{\rm eSZ}$ signals at high significance.  As we show below, the presence of foregrounds significantly changes the conclusion for $\mu$, but the outlook for the $y$ and $kT_{\rm eSZ}$ signals is still very positive.  

\begin{figure*}
\centering
\begin{subfigure}{.48\textwidth}
  \centering
  \includegraphics[width=\textwidth]{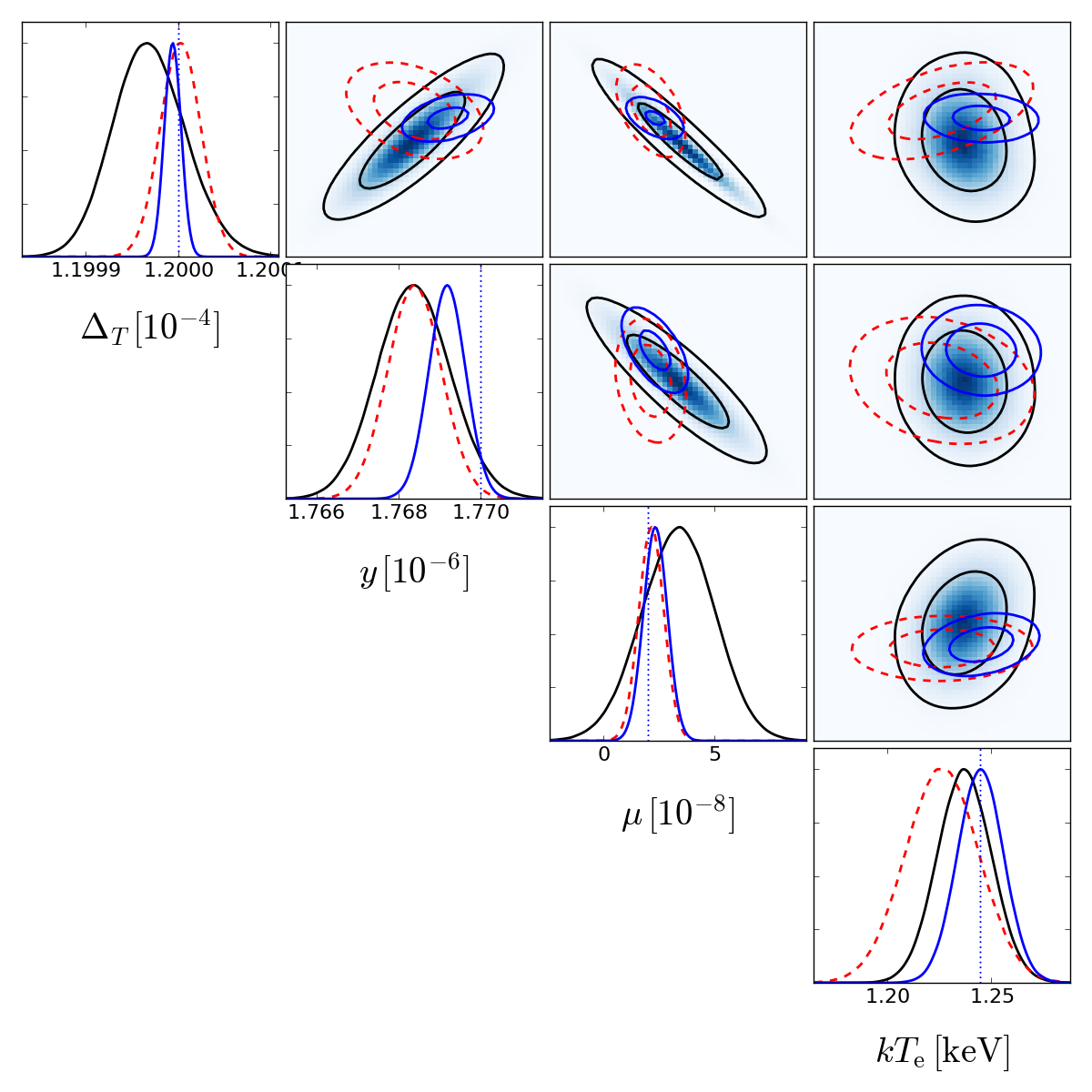}
\end{subfigure}%
\hspace{4mm}
\begin{subfigure}{.48\textwidth}
  \centering
  \includegraphics[width=\textwidth]{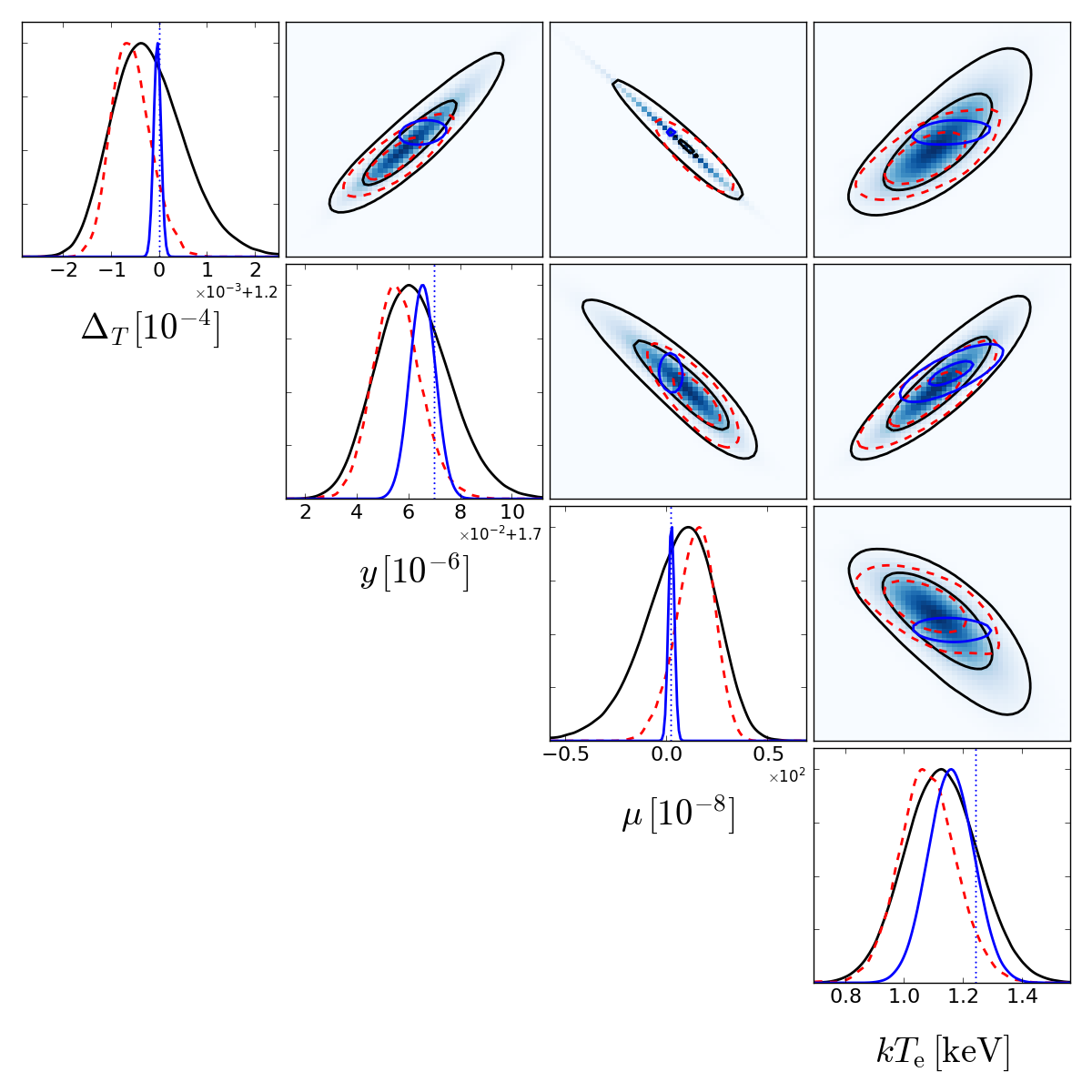}
\end{subfigure}%
  \caption{Comparison of the CMB spectral distortion parameter contours for varying foreground complexity. -- Left panel:  CMB-only (blue), CMB+Dust+CO (red), and CMB+Sync+FF+AME (black) parameter cases. Adding Dust+CO has a small effect on $\mu$, while adding Sync+FF+AME has a moderate effect on $kT_{\rm eSZ}$. -- Right panel: CMB+Dust+CIB+CO (blue), CMB+Sync+FF+Dust+CIB (red), and all foreground (black) parameter cases. The degradation of $\mu$ due to the foregrounds is more severe than that for the other parameters. The axis scales are different between the left and right panels and offsets are added.}
  \label{fig:triangleplot1}
\end{figure*}

\begin{table*}
\centering
\caption{Forecasts with foregrounds, using MCMC. All results are for the extended mission (86.4 months), except for the first column (12 months). The given numbers represent the average of the two-sided $1\sigma$ marginalized uncertainty on each parameter. The models for the extended mission are sorted using the errors on $y$ and $k\Te$. Values in parentheses are the detection significance (i.e., fiducial parameter value divided by 1$\sigma$ error). We assume a 10\% prior on the synchrotron amplitude and spectral index, $A_{\rm S}$ and $\alpha_{\rm S}$, to represent external datasets. This only has a noticeable effect for the 14 and 16 parameter cases. No band average is included, but this is found to have only a small effect.}
\begin{tabular}{lccccccccc}
\hline
\hline
 Sky Model
 & CMB 
 & CMB
 & Dust, CO 
 & Sync, FF,
 & Sync, FF,
 & Dust, CIB,  
 & Sync, FF, 
 & Sync, FF, AME
 \\[2pt]
 & (baseline)
 & 
 & 
 & AME
 & Dust
 & CO 
 & Dust, CIB
 & Dust, CIB, CO
 \\[2pt]
\# of parameters
& 4
& 4
& 8
& 9 
& 11
& 11
& 14 
& 16 
\\
\hline
\hline
$\sigma_{\Delta_T} [10^{-9}]$
& $2.3$ 
& $0.86$ 
& $2.2$ 
& $3.9$ 
& $ 9.7 $ 
& $5.3$ 
& $ 59 $ 
& $ 75$ 
\\[2pt]
$\sigma_{y} [10^{-9}]$
& $1.2$ ($1500\sigma$)
& $0.44$ ($4000\sigma$)
& $0.65$ ($2700\sigma$)
& $ 0.88 $ ($2000 \sigma$)
& $2.7$ ($660 \sigma$)
& $4.8$ ($370 \sigma$)
& $ 12 $ ($ 150 \sigma$)
& $ 14$ ($ 130 \sigma$)
\\[2pt]
$\sigma_{kT_{\rm eSZ}} [10^{-2}~\rm keV]$
& $2.9$ ($42 \sigma$)
& $1.1$ ($113 \sigma$)
& $1.8$ ($71\sigma$)
& $1.3$ ($96 \sigma$)
& $4.1$ ($30 \sigma$)
& $7.8$ ($16 \sigma$)
& $ 11$ ($ 11 \sigma$)
& $ 12 $ ($ 10\sigma$)
\\[2pt]
$\sigma_{\mu} [10^{-8}]$
& $1.4$ ($1.4 \sigma$)
& $0.53$ ($3.8 \sigma$)
& $0.55$ ($3.6 \sigma$)
& $1.7$ ($1.2\sigma$)
& $2.6$ ($ 0.76 \sigma$)
& $0.75$ ($2.7 \sigma$)
& $ 14 $ ($ 0.15 \sigma$)
& $18$ ($ 0.11 \sigma$)
\\[1pt]
\hline
\hline
\label{tab:MCMCfgforecast}
\end{tabular}
\end{table*}

\vspace{-0mm}
\section{Foreground-Marginalized Distortion Sensitivity Estimates} 
\label{sec:foregroundforecast}
We estimate the capability of a {\it PIXIE}-like experiment to detect the $\Delta_T$, $y$, $kT_{\rm eSZ}$, and $\mu$ spectral distortion parameters in the presence of the foregrounds described in Sect.~\ref{sec:foregrounds}. To understand the effect of each individual foreground on the distortion parameter forecast, we compare the effects of each component in Table~\ref{tab:MCMCfgforecast}. The forecasts assume an extended {\it PIXIE} mission and a 10\% prior on the synchrotron amplitude and index, $A_{\rm S}$ and $\alpha_{\rm S}$, unless stated otherwise. We find that in general the $\mu$ and $kT_{\rm eSZ}$ signals are the most obscured by foregrounds, which is expected since they are the faintest distortion signals.  $\Delta_T$ and $y$ are measured with high significance even in the worst cases. We therefore focus on the impact of the foregrounds on the $kT_{\rm eSZ}$ and $\mu$ spectral distortion parameters.  As a reference point, note that for the CMB-only extended mission, we found in the previous section that $kT_{\rm eSZ}$ is measured at 113$\sigma$ and $\mu$ at 3.8$\sigma$. Including all foreground parameters, this degrades to 10$\sigma$ for $kT_{\rm eSZ}$ and 0.11$\sigma$ for $\mu$ (Table~\ref{tab:MCMCfgforecast}).

We find that the $kT_{\rm eSZ}$ measurement is mainly affected by the high-frequency foregrounds of Galactic thermal dust and CIB, bringing the detection significance down to 16$\sigma$ when including these components. The integrated CO emission and dust alone have a more marginal effect (compare the 8 and 11 parameter dust cases in Table~\ref{tab:MCMCfgforecast}). This degradation is also illustrated in Fig.~\ref{fig:triangleplot1}, where we show the CMB distortion parameter posteriors for various sky models. It appears to be related to the fact that a superposition of modified blackbody spectra (thermal dust and CIB) produces a signal that mimics a Compton-$y$ distortion and relativistic correction in the Wien tail of the spectrum~\citep{Chluba2017}.

\begin{table*}
\centering
\caption{Errors on CMB parameters as a function of synchrotron parameter priors, using MCMC. These results assume an extended {\it PIXIE} mission and various priors (deduced from external data sets) on the synchrotron spectral index and amplitude, as indicated by the percentage values in the first row, respectively. In the final four columns, the $\mu$ parameter is not included in the data analysis (although it is present in the signal), yielding improved constraints on $kT_{\rm eSZ}$. For comparison, we also show the CMB-only (foreground-free) constraints ($4^{\rm th}$ and $8^{\rm th}$ column).}
\begin{tabular}{lccccccccc}
\hline
\hline
 Parameter
 & $1\%$  / --
 & $10\%$  / $10\%$
 & $1\%$ / $1\%$
 & CMB only
 & none (no $\mu$) 
 & $10\%$ / $10\%$ (no $\mu$)
 & $1\%$ / $1\%$ (no $\mu$)
 & CMB only
 \\
\hline
\hline
$\sigma_{\Delta_T} [10^{-9}]$
& $ 194 $ 
& $75$ 
& $18$ 
& $0.86$ 
& $ 17 $ 
& $ 4.4 $ 
& $ 3.7 $ 
& $ 0.42$ 
\\[2pt]
$\sigma_{y} [10^{-9}]$
& $ 32 $ ($ 55 \sigma$)
& $14$ ($130 \sigma$)
& $5.9$ ($300 \sigma$)
& $0.44$ ($4000\sigma$)
& $9.1 $ ($194\sigma$)
& $4.6$ ($380 \sigma$)
& $4.6$ ($390 \sigma$)
& $ 0.28 $ ($ 6200 \sigma$)
\\[2pt]
$\sigma_{kT_{\rm eSZ}} [10^{-2} \, \rm keV]$
& $ 23 $ ($ 5.5 \sigma$)
& $12$ ($ 10\sigma$)
& $8.6$ ($14 \sigma$)
& $1.1$ ($113 \sigma$)
& $ 12$ ($ 11 \sigma$)
& $7.9$ ($16 \sigma$)
& $7.6$ ($17 \sigma$)
& $ 1.0 $ ($120\sigma$)
\\[2pt]
$\sigma_{\mu} [10^{-8}]$
& $ 47 $ ($ 0.04 \sigma$)
& $18$ ($0.11\sigma$)
& $4.7$ ($0.43\sigma$)
& $0.53$ ($3.8 \sigma$)
& --
& --
& --
& --
\\[1pt]
\hline
\hline
\label{tab:forecastwpriors}
\end{tabular}
\end{table*}

\begin{table*}
\centering
\caption{Percent errors on foreground parameters, using MCMC. These results assume an extended {\it PIXIE} mission and various priors on the synchrotron spectral index and amplitude, as labeled in the first column. The average of the two-sided errors is quoted.  The recovered parameter posterior distributions for the final three cases (no $\mu$ in the analysis) are shown in Figure~\ref{fig:15p_no_mu_no_prior}. The synchrotron and free-free parameters are the least well constrained by {\it PIXIE}, suggesting that low-frequency ($\lesssim 15$~GHz) ground-based measurements could provide important complementary information.}
\begin{tabular}{ccccccccccccccccc}
\hline
\hline
Prior $\alpha_{\rm S}$ / $A_{\rm S}$
 & $A_{\rm S}$
 & $\alpha_{\rm S}$
 & $\omega_{\rm S}$
 & $A_{\rm FF}$
 & $A_{\rm AME}$
 & $A_{\rm d}$
 & $\beta_{\rm d}$
 & $T_{\rm d}$
 & $A_{\rm CIB}$
 & $\beta_{\rm CIB}$
 & $T_{\rm CIB}$
 & $A_{\rm CO}$
 \\
\hline
\hline
1\% / --
& $ 34.0 \%$ & $ 1.0 \%$ & $ 106.0 \%$ & $ 23.0 \%$ & $ 1.7 \%$ & $ 0.35 \%$ & $ 0.087 \%$ & $ 0.0051 \%$ & $ 1.2 \%$ & $ 0.32 \%$ & $ 0.1 \%$ & $ 0.33 \%$
\\ [1pt]
10\% / 10\%
& $ 9.6 \%$ & $ 9.3 \%$ & $ 52.0 \%$ & $ 7.3 \%$ & $ 0.9 \%$ & $ 0.18 \%$ & $ 0.051 \%$ & $ 0.0046 \%$ & $ 0.58 \%$ & $ 0.17 \%$ & $ 0.053 \%$ & $ 0.23 \%$
\\ [1pt]
1\% / 1\% 
& $ 0.99 \%$ & $ 1.0 \%$ & $ 5.5 \%$ & $ 1.1 \%$ & $ 0.77 \%$ & $ 0.13 \%$ & $ 0.04 \%$ & $ 0.0045 \%$ & $ 0.3 \%$ & $ 0.11 \%$ & $ 0.031 \%$ & $ 0.22 \%$
\\ [1pt]
\hline
\hline
none (no $\mu$)
& $ 33.0 \%$ & $ 29.0 \%$ & $ 93.0 \%$ & $ 8.9 \%$ & $ 1.3 \%$ & $ 0.18 \%$ & $ 0.048 \%$ & $ 0.0049 \%$ & $ 0.6 \%$ & $ 0.17 \%$ & $ 0.069 \%$ & $ 0.33 \%$
\\ [1pt]
10\% / 10\% (no $\mu$)
& $ 7.3 \%$ & $ 7.0 \%$ & $ 21.0 \%$ & $ 2.2 \%$ & $ 0.85 \%$ & $ 0.14 \%$ & $ 0.043 \%$ & $ 0.0046 \%$ & $ 0.35 \%$ & $ 0.12 \%$ & $ 0.029 \%$ & $ 0.21 \%$
\\ [1pt]
1\% / 1\% (no $\mu$)
& $ 0.95 \%$ & $ 0.95 \%$ & $ 5.1 \%$ & $ 0.47 \%$ & $ 0.61 \%$ & $ 0.12 \%$ & $ 0.038 \%$ & $ 0.0042 \%$ & $ 0.29 \%$ & $ 0.1 \%$ & $ 0.028 \%$ & $ 0.16 \%$
\\
\hline
\hline
\label{tab:priors}
\end{tabular}
\end{table*}

The $\mu$ distortion measurement is primarily obscured by the low-frequency synchrotron and free-free foregrounds (this will be further illustrated in Sect.~\ref{sec:priors}). The three synchrotron parameters are poorly constrained by {\it PIXIE} and significantly degrade the $\mu$ detection significance. The free-free spectrum is relatively flat in this frequency range and parameterized by only its amplitude, which is better measured than the synchrotron parameters by {\it PIXIE}. AME, the other low-frequency foreground, affects only a fairly narrow band and thus only has a small effect on $\mu$. These three foregrounds alone bring the $\mu$ detection significance down to 1.2$\sigma$ for an extended mission (see Table~\ref{tab:MCMCfgforecast} and Fig.~\ref{fig:triangleplot1}). 
Combining the four brightest components -- synchrotron, free-free, thermal dust, and CIB -- reduces the $kT_{\rm eSZ}$ detection significance to 11$\sigma$ and completely obscures the $\mu$ distortion (0.15$\sigma$). In the presence of all six foreground components, the $kT_{\rm eSZ}$ distortion is still detected at 10$\sigma$ significance (see Figure~\ref{fig:triangleplot1} and the last column of Table~\ref{tab:MCMCfgforecast}). However, the $\Lambda$CDM $\mu$-distortion seems to be out of reach without additional information.

For completeness, we list the forecasts for the baseline {\it PIXIE} mission with 12 months of spectral distortion mode integration time. With no priors (including, in this case, no priors on the synchrotron parameters) and all foreground components included, the $1\sigma$ uncertainties are: $\sigma_{\Delta T}=\pot{6.9}{-7}$, $\sigma_{y}=\pot{1.2}{-7}$, $\sigma_{kT_{\rm eSZ}}=0.9$~keV, and $\sigma_{\mu}=\pot{1.5}{-6}$. Comparing to the $1\sigma$ limits of {\it COBE/FIRAS}, $\sigma_{\Delta T}\simeq \pot{2.0}{-4}$, $\sigma_{y}\simeq\pot{7.5}{-6}$ and $\sigma_{\mu}\simeq\pot{4.5}{-5}$, shows that with 12 months in spectral distortion mode {\it PIXIE} will improve the parameter constraints by more than a factor of $\simeq 30$. However, this comparison is not precisely valid, as the {\it COBE/FIRAS} limits are derived on combinations of two parameters only ($\sigma_{\Delta T}$ and $y$ or $\mu$). For example, omitting $\mu$ and $kT_{\rm eSZ}$ we find the baseline sensitivities $\sigma_{\Delta T}=\pot{4.1}{-8}$ and $\sigma_{y}=\pot{8.7}{-9}$, which highlights the large improvement in the raw sensitivity ($\gtrsim 1000$ times better than {\it COBE/FIRAS}; see Fig.~\ref{fig:sdsignals}.).\footnote{We show the raw sensitivity of the {\it COBE/FIRAS} mission in Fig.~\ref{fig:sdsignals}, but the cosmological parameter constraints quoted in the text include additional degradation due to systematic errors. Furthermore, as mentioned in Sect.~\ref{sec:Intro}, the {\it COBE/FIRAS} analysis methodology differs significantly from ours. In particular, that analysis relies entirely on spatial information to separate Galactic foregrounds from the extragalactic monopole, while ours relies entirely on spectral information to separate different components at the level of the monopole SED.  An optimal analysis would combine both sets of information, but this requires the simulation of detailed sky maps at each {\it PIXIE} frequency.}

Assuming 10\% priors on the synchrotron index and amplitude and the baseline mission sensitivity gives: $\sigma_{\Delta T}=\pot{9.6}{-8}$, $\sigma_{y}=\pot{2.1}{-8}$, $\sigma_{kT_{\rm eSZ}}=0.25$~keV, and $\sigma_{\mu}=\pot{2.3}{-7}$. The priors carry a significant amount of information about the low-frequency foregrounds. In the framework of this analysis, the biggest gains on CMB parameters come from better constraining these foregrounds, in particular the synchrotron emission. A side effect of including external priors is that they necessarily reduce the efficiency of increasing the mission sensitivity in a Fisher analysis. The extended mission has $\approx\sqrt{86.4/12}\approx 2.68$ times better sensitivity than the baseline mission, but we see only a factor of $\approx 1.3$ improvement in the CMB parameter constraints due to the external priors dominating the information on the synchrotron SED. When comparing the baseline and extended mission without priors, we find an improvement of almost exactly 2.68, but the constraints are of course better with the external priors applied, as seen in Figures~\ref{fig:fg_fminfstep}, ~\ref{fig:fgkt}, and~\ref{fig:sensitivity} (discussed in detail below).

\vspace{-0mm}
\subsection{Foreground Model Assumptions}
\label{sec:foreground_model}
We consider the effects of varying the foreground models and parameter values on the spectral distortion forecast using the Fisher method. First, we vary each foreground amplitude parameter by up to a factor of 5 and find very little change in the projected spectral distortion uncertainties. Next, we find that the forecasts are still accurate when varying the spectral indices or component temperatures by up to $\approx 20\%$. Further modification of the spectral shape parameters, in particular the synchrotron spectral index and curvature, can noticeably change the forecast estimates, but these modifications are not consistent with current observations \citep[e.g.,][]{wmap9results, planck2015Xforegrounds} which indicate that $\alpha_{\rm sync}$ only varies at the $\simeq 5-10\%$ level across the sky. 

We also consider simplifying the synchrotron model to a two-parameter power law, which improves the detection significance by about a factor of 2 on $\mu$ and by 1.3 on $kT_{\rm eSZ}$. This is expected, since we saw previously that low-frequency foregrounds mainly degrade the detection significance of $\mu$ (e.g., Fig.~\ref{fig:triangleplot1}). However, this scenario is again unrealistic, as the curvature of the synchrotron spectrum is an inevitable result of the average of the synchrotron emission over the sky and along the line-of-sight. Spatial information on $\alpha_{\rm sync}$ could be used to further constrain $\omega_{\rm sync}$, but the effects of line-of-sight and beam averaging will not be separable in this way. In addition, due to the rather low angular resolution of {\it PIXIE}, a combination with other experiments might be required in this case, so that we leave this aspect for future explorations.

Removing the CIB emission entirely results in the largest (factor of $\simeq 5$) improvement in the detection of all CMB parameters, but this is also unrealistic. Rather, we expect a more complicated model for the dust components to be required, which directly handles and models the information from spatial variations of the SED parameters. 
Allowing the peak frequency of the spinning dust SED to be a free parameter negligibly affects the $\mu$ and $kT_{\rm eSZ}$ detection significance when assuming the 10\% prior on $A_{\rm sync}$ and $\alpha_{\rm sync}$. In fact with this synchrotron prior, the entire spinning dust SED provides only a marginal ($< 20\%$) reduction in CMB parameter detections. Even when relaxing the synchrotron priors, the spinning dust affects the CMB parameters' detection significance by less than a factor of 2. Overall, we find that the forecast is robust to moderate variations in the assumed foreground model and parameters. Rather than the specific amplitude of the signals, the shapes and covariance with the distortion parameters is most important in driving the CMB parameter limits.

\vspace{-0mm}
\subsection{Addition of External Data Using Priors}
\label{sec:priors}
We examine the use of external information in the form of {\it a priori} knowledge of the foreground SED parameters. The biggest improvements can be expected for the low-frequency foreground parameters, as the high-frequency components generally are found to be constrained with high precision ($\lesssim 1\%$) due to the large number of high-frequency channels in FTS concepts. We thus compare results using combinations of 10\% and 1\% priors on the synchrotron amplitude and spectral index in Tables~\ref{tab:forecastwpriors} and~\ref{tab:priors}. This is meant to mimic information from future ground-based experiments similar to C-BASS and QUIJOTE, possibly with extended capabilities related to absolute calibration, or when making use of extra spatial information in the analysis.

Focusing on Table~\ref{tab:forecastwpriors}, the $\Lambda$CDM $\mu$ distortion is still not detectable even with tightened priors, but the $kT_{\rm eSZ}$ detection significance could be improved to 14$\sigma$ when imposing $1\%$ priors on $A_{\rm sync}$ and $\alpha_{\rm sync}$. In this case, an upper limit of $|\mu|< 9.4\times 10^{-8}$ (95\% c.l.) could be achieved. Comparing this with the CMB-only constraints reveals that foregrounds introduce about one order of magnitude degradation of the constraint. 
To detect the $\Lambda$CDM $\mu$ distortion at $2\sigma$ requires a $\simeq 0.1\%$ prior on the synchrotron amplitude, index, curvature, and the free-free amplitude. This is not met by {\it PIXIE} alone, but could possibly be achieved by adding constraints from ground-based observations at lower frequencies. In particular the steepness of the synchrotron SED might help in this respect, with increasing leverage as lower frequencies are targeted. Performing similar measurements with a space mission will be very challenging due to constraints on the size of the instrument.

Nevertheless, {\it PIXIE} could significantly improve the existing limit from {\it COBE/FIRAS}, placing tight constraints on the amplitude of the small-scale scalar power spectrum, $A_{\rm s}$, around wavenumber $k\simeq 740\,\Mpc^{-1}$, corresponding to \citep[cf.,][]{Chluba2013PCA}
\beal
A_{\rm s}(k \simeq 740\,\Mpc^{-1})< \pot{2.8}{-8}\,\left[\frac{|\mu|}{\pot{3.6}{-7}}\right]\, \text{(95\% c.l.)},
\end{align}
when assuming a scale-invariant (spectral index $n_{\rm s}=1$) small-scale power spectrum. Here, $|\mu|$ is the $2\sigma$ upper limit on the chemical potential. This would already rule out many alternative early-universe models with enhanced small-scale power \citep{Chluba2012inflaton, Clesse2014}. Assuming $1\%$ priors on $A_{\rm sync}$ and $\alpha_{\rm sync}$, one could obtain $A_{\rm s}(k \simeq 740\,\Mpc^{-1})< \pot{7.3}{-9}\text{(95\% c.l.)}$, bringing us closer to the value obtained from CMB anisotropy observations, $A_{\rm s}(k\simeq 0.05\,\Mpc^{-1})\simeq \pot{2.2}{-9}$ \citep{Planck2015params} at much larger scales.

\begin{figure}
\centering
\includegraphics[width=\columnwidth]{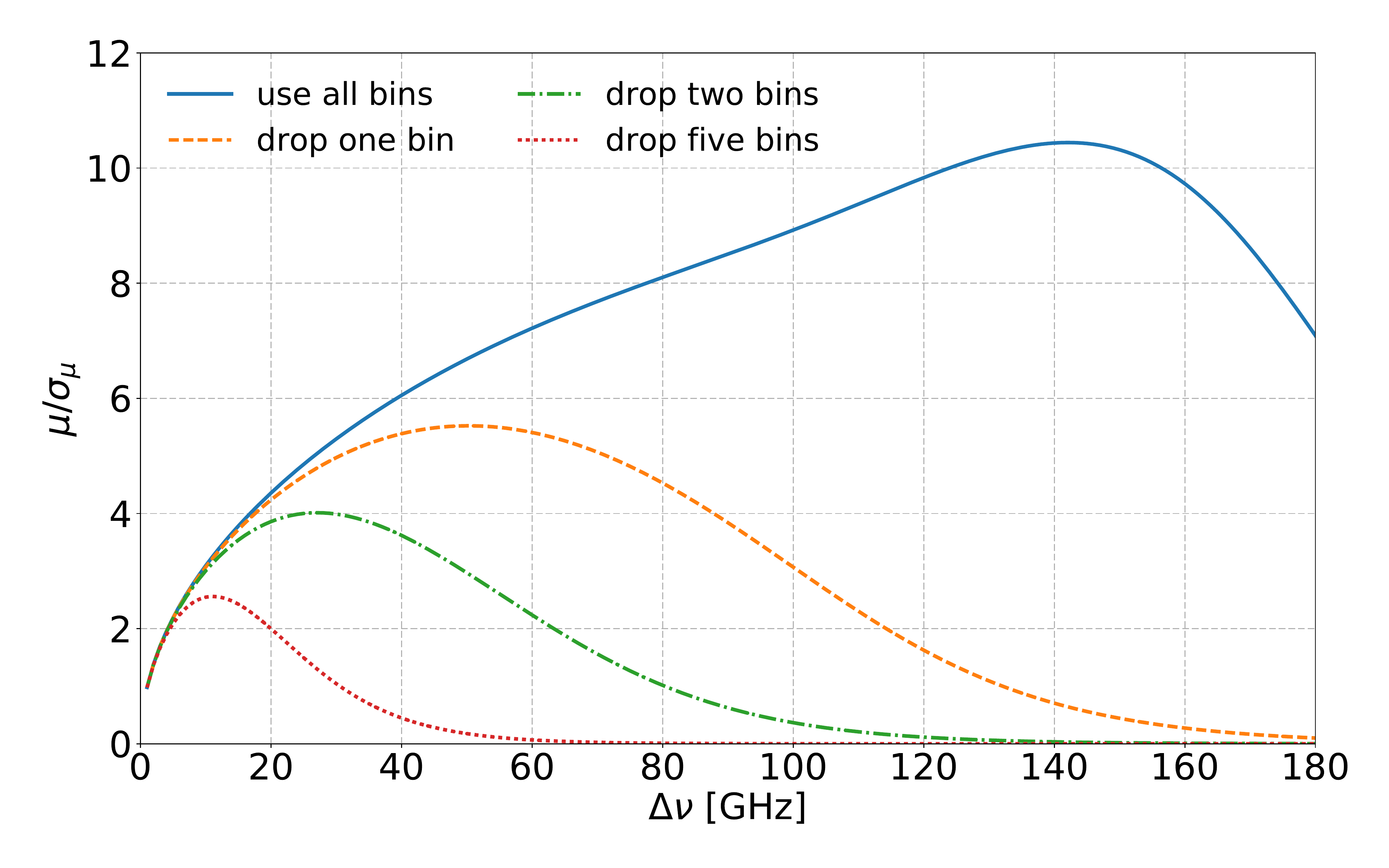} 
\\
\includegraphics[width=\columnwidth]{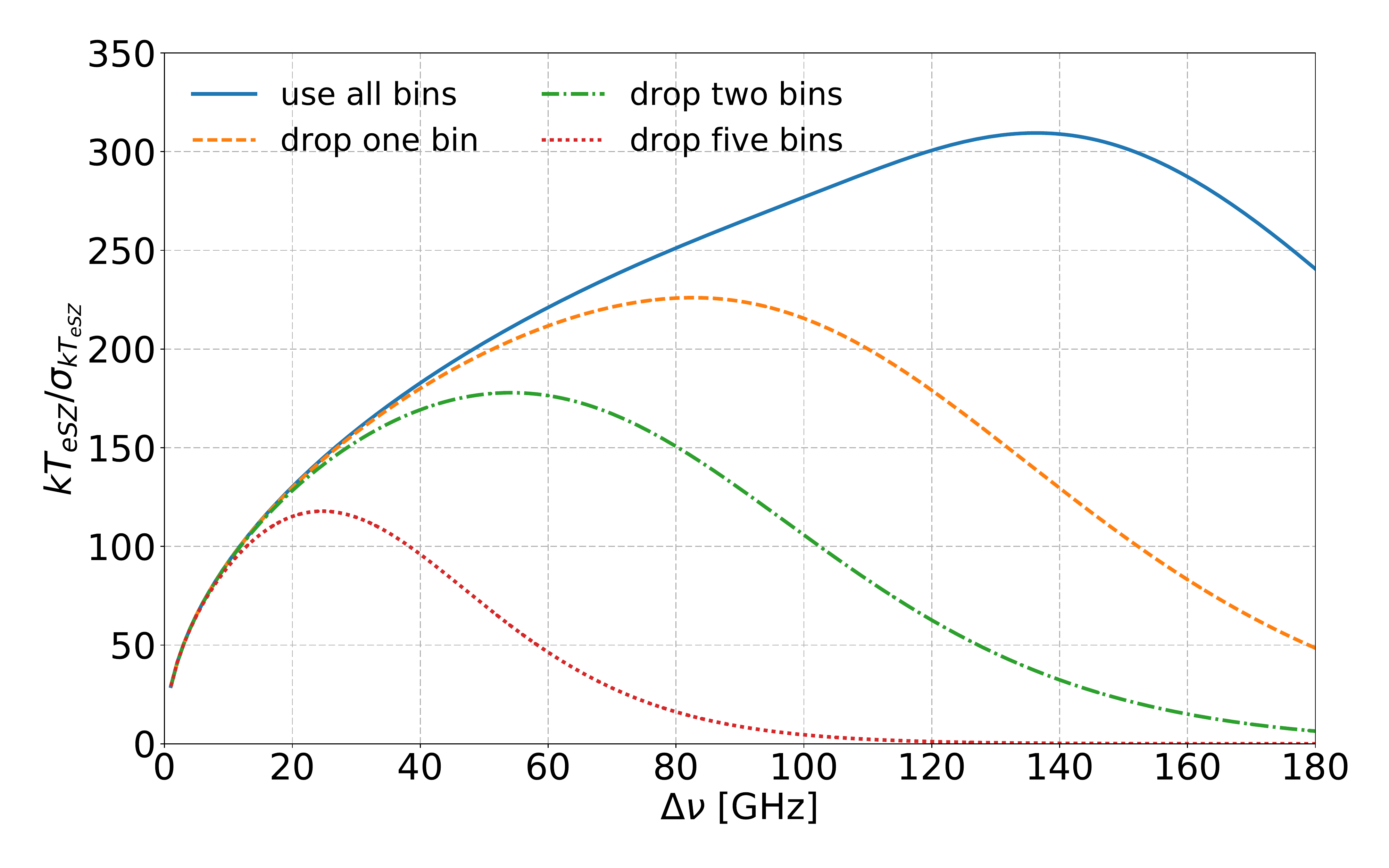}
  \caption{Estimated CMB-only (i.e., no foregrounds; extended mission) detection significance for the $\Lambda$CDM $\mu$ (upper panel) and $kT_{\rm eSZ}$ (lower panel) signals as a function of the frequency resolution, $\Delta \nu$. The different curves show the effect of dropping a varying number of the lowest-frequency channels to mimic systematic-related channel degradation.}
 \label{fig:cmb_fminfstep}
\end{figure}

As mentioned above, the biggest issue for the $\mu$ detection indeed lies in the synchrotron emission. Table~\ref{tab:priors} shows the expected uncertainties on foreground parameters with various assumed synchrotron SED priors. The thermal dust, CIB, and CO parameters are all measured to $<1\%$, while the low-frequency foregrounds have much larger uncertainties. This implies that the largest gains in terms of CMB distortion science can be expected by improving ground-based measurements at low frequencies. These measurements will also be required to complement CMB $B$-mode experiments, aiming at detection of a tensor-to-scalar ratio $r< 10^{-3}$.

Generally, we also find that imposing a prior on $\alpha_{\rm sync}$ alone does not significantly improve the results. For example, we find the distortion constraints with $10\%$ and $1\%$ prior on $\alpha_{\rm sync}$ to be practically the same (the $1\%$ prior on $\alpha_{\rm sync}$ case is shown in Table~\ref{tab:forecastwpriors}). This is due to the strongly non-Gaussian posteriors of the model parameters (see Fig.~\ref{fig:15p_no_mu_no_prior} for the cases without $\mu$) and adding a $10\%$ prior on $A_{\rm sync}$ immediately improves the CMB distortion constraints by a factor $\simeq 2$. This means that it will be crucial to obtain additional low-frequency constraints on the absolute sky-intensity, while simple differential measurements will only help in constraining the spatially-varying contributions to $\omega_{\rm sync}$.

Further improvements for $kT_{\rm eSZ}$ are seen when $\mu$ is excluded from the parameter analysis (although the signal is still present in the sky model).  In this case, the $kT_{\rm eSZ}$ detection significance can reach $17\sigma$ (see Table~\ref{tab:priors}). The addition of $10\%$ priors on $A_{\rm sync}$ and $\alpha_{\rm sync}$ are sufficient to achieve this. This can also be seen in Fig.~\ref{fig:15p_no_mu_no_prior} and stems from the highly non-Gaussian tails of the $kT_{\rm eSZ}$ posterior without external prior.
We also find small biases in the deduced distortion parameters (see Fig.~\ref{fig:15p_no_mu_no_prior}). For example, for the case with $1\%$ priors on $A_{\rm sync}$ and $\alpha_{\rm sync}$, we obtain $y=\pot{(1.7676 \pm 0.0045 )}{-6}$ and $kT_{\rm eSZ}=(1.182\pm 0.075)\,{\rm keV}$, corresponding to marginal biases of $\simeq -0.5\,\sigma$ and $\simeq -0.8\,\sigma$, respectively. These can usually be neglected. CMB spectral distortion measurements are thus expected to yield robust constraints on $kT_{\rm eSZ}$.

\vspace{-0mm}
\subsection{Optimal Mission Configuration Search}
\label{sec:optimal}
In an effort to optimize the instrument configuration in the presence of foregrounds, we study the effect of varying the mission parameters, such as sensitivity, frequency resolution, and frequency coverage, all assuming an FTS concept. Aside from extending the mission duration, the overall sensitivity can only be increased by increasing the aperture and detector array (to increase the \'etendue), as concepts like {\it PIXIE} are already photon-noise-limited. The FTS mirror stroke controls the frequency resolution and the physical size of the detector array limits the lowest frequency channels. This implies that changing the above experimental parameters is a strong cost driver, and complementarity between different experimental concepts needs to be explored.

We characterize the mission in terms of the detection significance for the CMB spectral distortion fiducial $\Lambda$CDM $\mu$ and $kT_{\rm eSZ}$ parameters ($y$ and $\Delta_T$ are detected at high significance in all scenarios and will not be further highlighted). We assume the extended mission and consider varying the priors on the synchrotron amplitude and spectral index, $A_{\rm S}$ and $\alpha_{\rm S}$. The lower edge of the lowest frequency channel is set to $\nu_{\rm min}\simeq 7.5\,\GHz$ (similar to {\it PIXIE}), determined by the physical dimension of the instrument. We assume an otherwise ideal instrument with a top-hat frequency response\footnote{For large $\Delta\nu$, the band average becomes very important.} and, in terms of systematics, consider white noise only. 

\subsubsection{Optimal setup without foregrounds}
In Figure~\ref{fig:cmb_fminfstep}, we show the estimated CMB-only detection significance for the $\Lambda$CDM $\mu$ and $kT_{\rm eSZ}$ signals as a function of the frequency resolution, $\Delta \nu$, which in principle can be varied in-flight by adjusting the mirror stroke. The sensitivity per channel scales as $\simeq \Delta \nu/15\,\GHz$ (i.e., wider frequency bins have higher sensitivity).
The lowest frequency channels are susceptible to instrument-related systematic errors, so we also consider forecasts in which we drop a fixed number of the lowest frequency bins. 
When ignoring foreground contamination, the optimal configuration for measuring the $\mu$ distortion is $\Delta \nu\simeq 142$~GHz when all channels are included, yielding a $10.4\sigma$ detection of the standard $\Lambda$CDM value, $\mu\simeq \pot{2}{-8}$. This drops steeply to $\simeq 5.5\sigma,\, 4.0\sigma,\, 2.6\sigma$ for optimal resolutions $\Delta \nu\simeq 50\,\GHz,\, 27\,\GHz,\, 11\,\GHz$, if the lowest one, two, or five frequency channels cannot be used, respectively.

A similar trend is found for the optimal configuration aiming to detect $kT_{\rm eSZ}$ (lower panel, Fig.~\ref{fig:cmb_fminfstep}), with the optimal resolution being $\Delta \nu\simeq 135$~GHz when all channels are included in the analysis, giving a $>300\sigma$ detection of the signal. This degrades to $\simeq 226\sigma,\, 178\sigma,\, 118\sigma$ for optimal resolutions $\Delta \nu\simeq 83\,\GHz,\, 54\,\GHz,\, 25\,\GHz$, if the lowest one, two or five frequency channels cannot be used, respectively.

The error is mainly driven by the competition between sensitivity per channel ($\leftrightarrow$ frequency bin size) and frequency coverage ($\leftrightarrow$ lowest frequency bin) to allow the separation of the distortion parameters, with $kT_{\rm eSZ}$ and $\mu$  usually most strongly correlated. In particular, the sensitivity to $\mu$ drops when the design is near a regime in which no independent frequency channel is present below the null of the $\mu$-distortion signal at $\nu\simeq 130\,\GHz$. For instance, assuming all channels can be included, one would expect a configuration with one bin between $7.5\,\GHz$ and $130\,\GHz$, giving $\Delta\nu\simeq 120\GHz$, to be roughly optimal. Dropping the lowest frequency bin, one would expect a configuration with two bins between $7.5\,\GHz$ and $130\,\GHz$, giving $\Delta\nu\simeq 120\GHz/2\simeq 60\,\GHz$, to be optimal, and so on. These numbers are in good agreement with the true optimal frequency bin widths found above.

\begin{figure}
\centering
\includegraphics[width=\columnwidth]{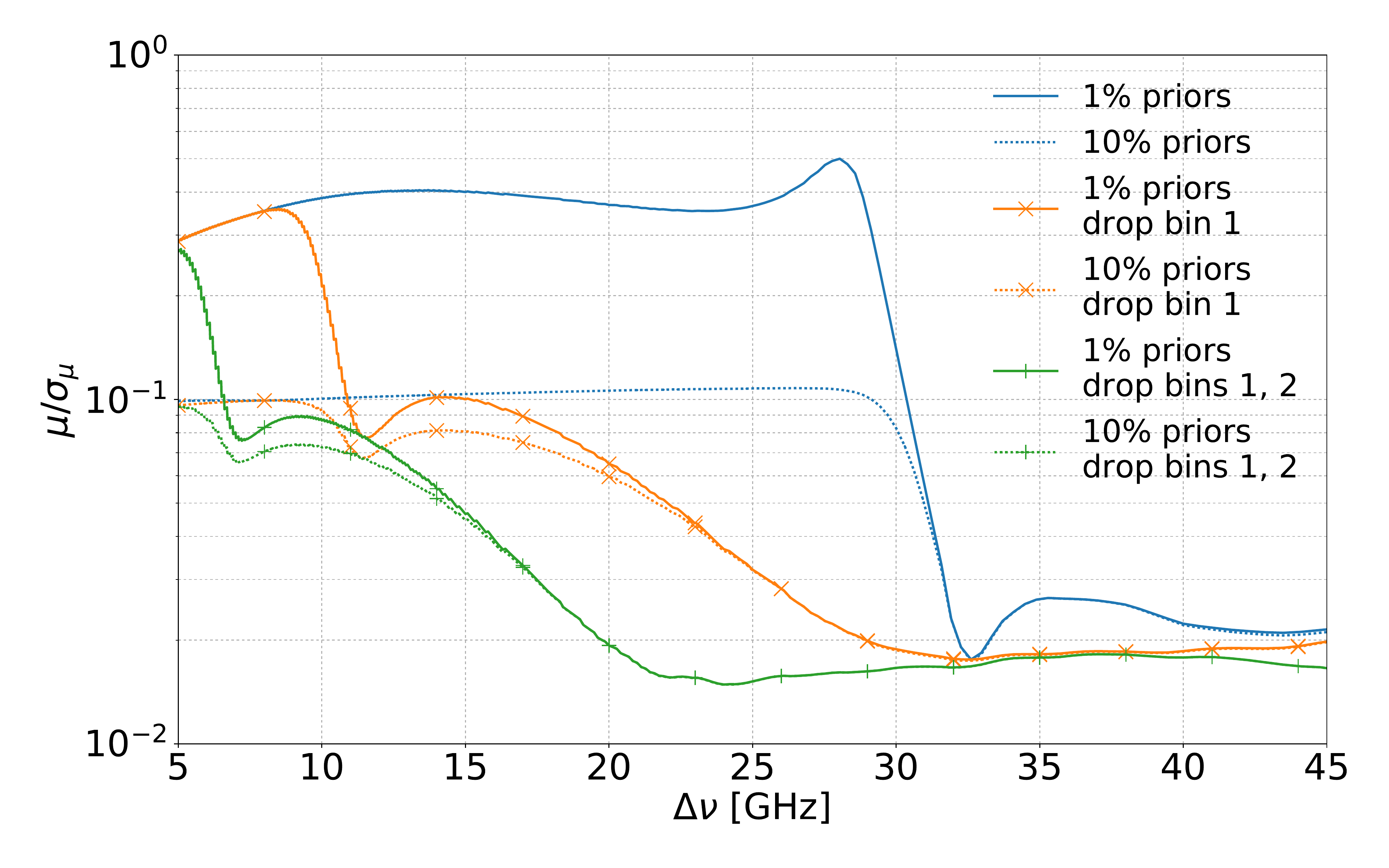}
\caption{Estimated detection significance (foregrounds included; extended mission) for the $\Lambda$CDM $\mu$-distortion signal as a function of the frequency resolution, $\Delta \nu$ (note the logarithmic scale on the vertical axis). The different curves show the effect of dropping the lowest-frequency channels and changing the priors on $A_{\rm sync}$ and $\alpha_{\rm sync}$.}
 \label{fig:fg_fminfstep}
\end{figure}
\subsubsection{Optimal setup with foregrounds}
\begin{figure}
\includegraphics[width=\columnwidth]{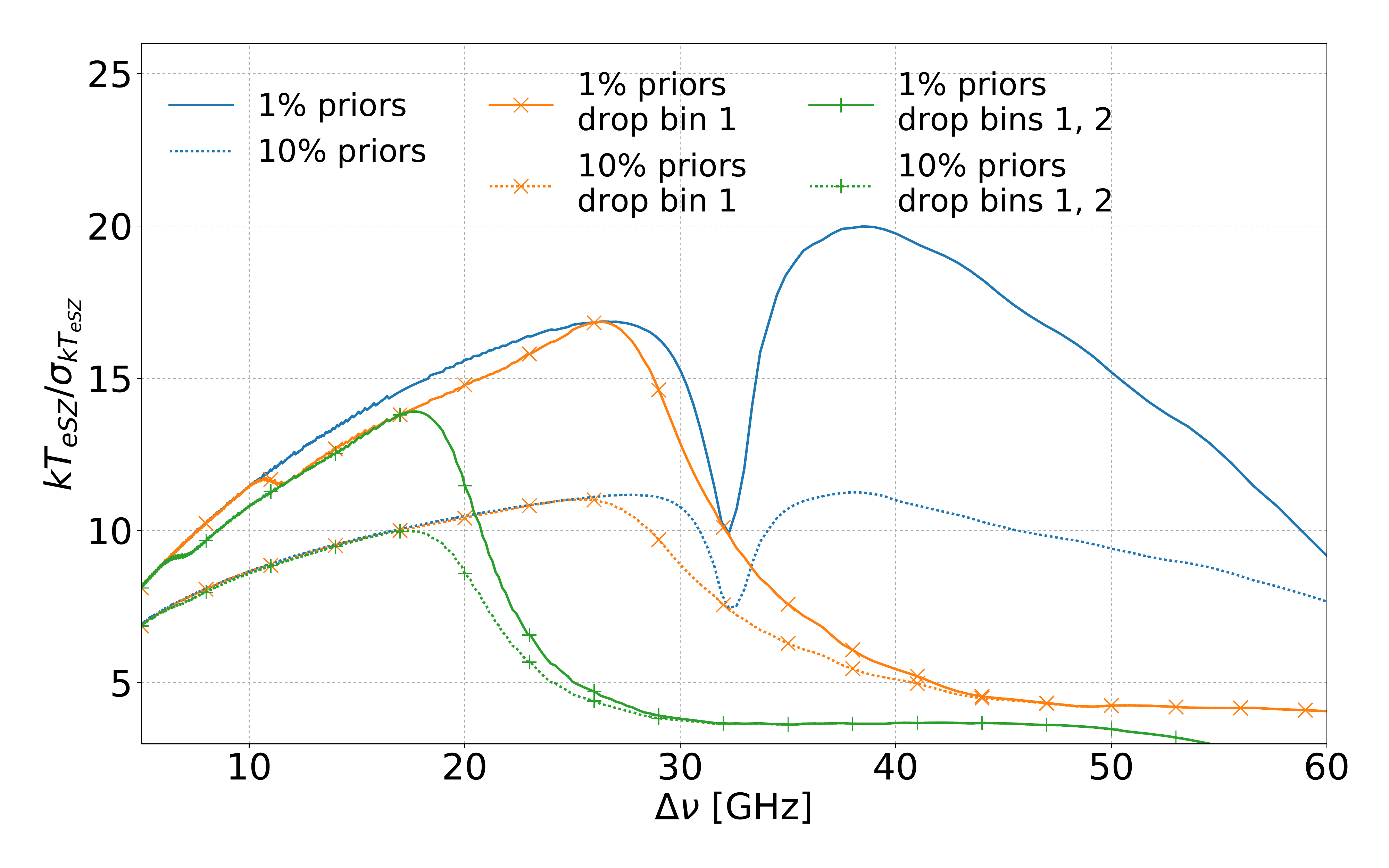} 
\\
 \includegraphics[width=\columnwidth]{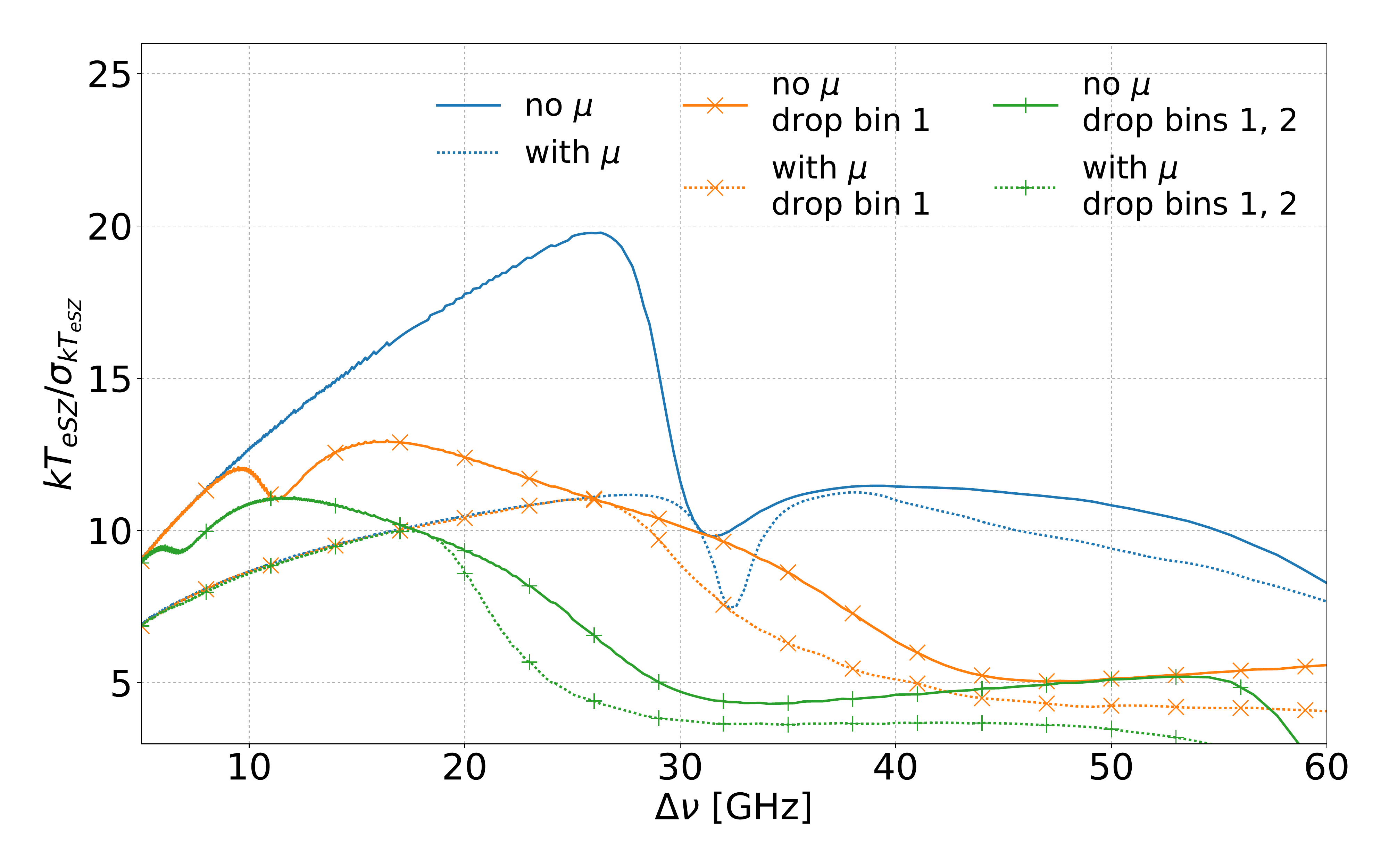}
  \caption{Estimated detection significance (foregrounds included; extended mission) for the $\Lambda$CDM $kT_{\rm eSZ}$ signal as a function of the frequency resolution, $\Delta \nu$. The different curves show the effect of omitting the lowest-frequency channels and changing the priors on $A_{\rm sync}$ and $\alpha_{\rm sync}$. The upper panel illustrates the results when $\mu$ is included in the analysis, while in the lower panel the cases for 10\% priors are compared with and without $\mu$ included.}
  \label{fig:fgkt}
\end{figure}

\begin{figure}
	\includegraphics[width=\columnwidth]{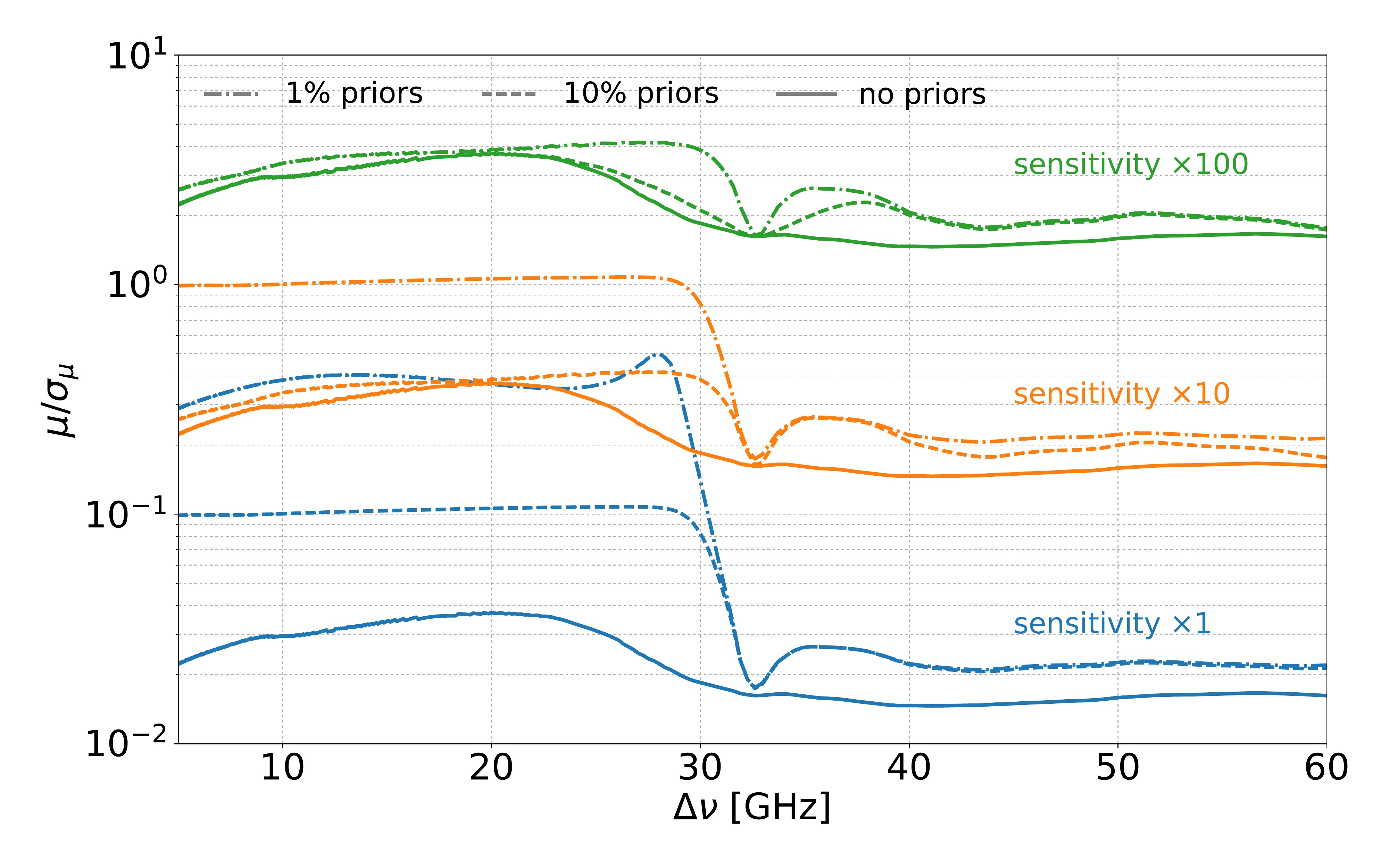}
  \caption{Estimated detection significance (foregrounds included; extended mission) for the $\Lambda$CDM $\mu$-distortion signal as a function of the frequency resolution, $\Delta \nu$, and for varying priors on $A_{\rm sync}$ and $\alpha_{\rm sync}$ and increasing  the overall mission sensitivity (note the logarithmic scale on the vertical axis).}
  \label{fig:sensitivity}
\end{figure}

When including foreground contamination, the picture changes significantly. Focusing on the detection significance for $\mu$ (Fig.~\ref{fig:fg_fminfstep}), we see that when including all channels the optimal frequency resolution is $\Delta \nu\lesssim 27\,\GHz$, independent of the chosen prior on the synchrotron parameters (blue curves in Fig.~\ref{fig:fg_fminfstep}). The sensitivity remains rather constant in this regime, since most of the information is already delivered by including external data as represented by the $10\%$ or $1\%$ priors on $A_{\rm sync}$ and $\alpha_{\rm sync}$. A sharp drop in the $\mu$-sensitivity is found around $\Delta \nu\simeq 30\,\GHz$. This is roughly where in our model the transition between low- and high-frequency foreground components occurs (see Fig.~\ref{fig:fgsignals}), driving the trade-off in the frequency resolution toward lower frequencies. For $\Delta \nu\simeq 30\,\GHz$ all of the low-frequency foreground information is contained in one channel, which limits the ability of such a setup to separate individual components. This feature is also seen in Figures~\ref{fig:fgkt} and~\ref{fig:sensitivity} (discussed below).

The sensitivity, even for the extended mission, is not sufficient to detect the $\Lambda$CDM $\mu$ distortion, but greatly improved limits of $|\mu|\lesssim \pot{\rm few}{-7}$ are within reach. The increase in detection significance at lower frequencies is due primarily to better constraining the synchrotron and free-free SEDs. Dropping the lowest frequency channels further pushes the optimal frequency resolution to $\Delta \nu\lesssim 15$ GHz. This statement is relatively independent of the assumed synchrotron priors and indicates that the $\mu$-distortion sensitivity of {\it PIXIE} is relatively robust with respect to the inclusion of the lowest FTS channels. However, modest improvements are seen when choosing $\Delta \nu\lesssim 10\,\GHz$.

Figure~\ref{fig:fgkt} shows the detection significance for $kT_{\rm eSZ}$ when varying experimental parameters as above. We also show the constraints when $\mu$ is excluded from the analysis (lower panel).
For the relativistic tSZ parameter, the optimal frequency resolution is $\Delta \nu\simeq 27\,\GHz$ or $\simeq 37\,\GHz$ when all channels are included in the analysis. For $\Delta \nu\simeq 27\,\GHz$, the sensitivity to $\mu$ is also optimized. Although this is not the default resolution of {\it PIXIE}, the improvement in the distortion sensitivity over $\Delta \nu\simeq 15\,\GHz$ is only $\simeq 15\%$. $\Delta \nu\simeq 27\,\GHz$ is also optimal when dropping the lowest frequency channel, but in this case drops off rapidly for $\Delta \nu\gtrsim 30\,\GHz$. When dropping the lowest two frequency channels, the optimal frequency resolution is $\Delta \nu\simeq 17\,\GHz$, which is very close to the default setting of {\it PIXIE}. In this case, the sensitivity to $kT_{\rm eSZ}$ furthermore drops strongly for $\Delta \nu\gtrsim 20\,\GHz$.

Ignoring $\mu$ in the parameter analysis generally yields improved constraints on $kT_{\rm eSZ}$ (lower panel, Fig.~\ref{fig:fgkt}). In particular, gains are seen for $\Delta \nu\lesssim 30\,\GHz$, as low-frequency information can be used to improve the constraint without competition from $\mu$. Again, $\Delta \nu \simeq 15\,\GHz$ is found to optimize the overall trade-offs when the lowest frequency channel is omitted, and $\Delta\nu\simeq 11\,\GHz$ is optimal when dropping the lowest two channels.

We also study the dependence of the $\mu$ constraints on the overall sensitivity of the instrument, which could be modified by increasing the total aperture and detector array (e.g., by replicating the instrument) or further extending the mission duration.
As expected, this improves the spectral distortion measurements (Fig.~\ref{fig:sensitivity}). In particular, increasing the mission sensitivity by a factor of 100 enables a $\simeq 3-4\sigma$ detection of the expected $\Lambda$CDM $\mu$ distortion, $\mu\simeq \pot{2}{-8}$, for the default setup. This implies an \'etendue 10,000 times greater than that of {\it PIXIE}, which is prohibitively expensive with current technology. Combining {\it PIXIE} with 1\% synchrotron index and amplitude priors performs as well as an experiment with 10 times the sensitivity and no priors when focusing on the $\mu$ constraints, for $\Delta\nu\lesssim 30\,\GHz$. This emphasizes the impact external synchrotron datasets, possibly also exploiting spatial information, could have when combined with {\it PIXIE}. 
 
Our analysis shows that ultimately the biggest hurdle to measuring $\mu$ comes down to the low-frequency foregrounds, in particular the synchrotron and free-free emission. A dedicated ground-based, low-frequency instrument with spatial resolution at least as good as {\it PIXIE} will be essential in constraining these foregrounds. Alternatively, one could think about adding a low-frequency instrument not based on the FTS concept to the payload. A detailed analysis of these ideas is left to future work, but a rough estimate with the Fisher method seems to indicate that tens of spectral channels between $\simeq 1$ and 30~GHz with sensitivity equal to or better than the baseline {\it PIXIE} mission would be required. This leads to $<1\%$ constraints on the synchrotron, free-free, and spinning dust SEDs and could yield $|\mu| < 2.0\times 10^{-8}$ (95\% c.l.).

\vspace{-3mm}
\section{Conclusions}
\label{sec:conclusions}
Measurements of CMB spectral distortions will shed new light on physics in both the early ($\leftrightarrow \mu$ distortion) and late ($\leftrightarrow y$ and $kT_{\rm eSZ}$ signals) periods of cosmic history. This will open a new era in CMB cosmology, with clear distortion signals awaiting us.
However, detecting spectral distortions will require extreme precision and control of systematics over large bandwidths. 
Our analysis shows that foregrounds strongly affect the expected results and demand dedicated observations and experimental designs, particularly with improved sensitivity at low frequencies ($\nu\lesssim 15-30\,\GHz$). 

We considered foreground models motivated by available CMB observations to forecast the capability of {\it PIXIE} or other future missions to detect spectral distortions, focusing on FTS concepts. We find that {\it PIXIE} has the capability to measure the CMB temperature to $\simeq$ nK precision and to detect the Compton-$y$ and relativistic tSZ distortions at high significance (see Table~\ref{tab:forecastwpriors}). 
With conservative assumptions about extra information at low frequencies from external data, we expect detections at 194$\sigma$ and 11$\sigma$ for $y$ and $kT_{\rm eSZ}$, respectively.
We emphasize that $kT_{\rm eSZ}$ is detected at above $5\sigma$ with no modification of the {\it PIXIE} mission and no external data. The $kT_{\rm eSZ}$ detection significance is increased to $>11\sigma$ when $\mu$ is ignored in the parameter analysis (shown in the last 3 columns of Table~\ref{tab:forecastwpriors} and in Figure~\ref{fig:15p_no_mu_no_prior}).
These measurements would provide new constraints on models of baryonic structure formation, thus providing novel information about astrophysical feedback mechanisms \citep{Hill2015, Battaglia2016}.

Due to its many high-frequency channels, {\it PIXIE} will provide the best measurements of thermal dust and CIB emission to date (see Table~\ref{tab:priors}). These sub-percent absolute measurements will provide invaluable information for the modeling of CMB foregrounds relevant to $B$-mode searches from the ground. They will furthermore allow improvements in the channel inter-calibration, potentially allowing us to reach sensitivities required to extract resonant scattering signals caused by atomic species~\citep[e.g.,][]{Kaustuv2004, Jose2005}. They will also greatly advance our understanding of Galactic dust properties and physics, providing invaluable absolutely calibrated maps in many bands.

The fiducial $\Lambda$CDM $\mu$ distortion ($\mu\simeq \pot{2}{-8}$) is unlikely to be detected in the presence of known foregrounds without better sensitivity or additional high-fidelity datasets that constrain the amplitude and shape of low-frequency foregrounds. Nevertheless, {\it PIXIE} could improve upon the existing limit from {\it COBE/FIRAS} by a factor of 250, yielding $|\mu| < 3.6\times 10^{-7}$ (95\% c.l.) for conservative assumptions about available external data. This would place tight constraints on the amplitude of the small-scale scalar power spectrum at wavenumber $k\simeq 740\,\Mpc^{-1}$ (Sect.~\ref{sec:priors}) and would rule out currently allowed parameter space for long-lived decaying particles with lifetimes $\simeq 10^{6}-10^{10}\,{\rm s}$ \citep[e.g., see][]{Chluba2013PCA}.

We find {\it PIXIE}'s frequency resolution of $\Delta \nu\simeq 15\,\GHz$ to be close to optimal, although for $\mu$ slight improvements could be expected when using $\Delta \nu\simeq 10\,\GHz$, depending on the quality of the low-frequency channels (Sect.~\ref{sec:optimal}). A similar choice seems optimal for $kT_{\rm eSZ}$, in particular if the lowest frequency channels cannot be used in the data analysis due to systematic errors (see Fig.~\ref{fig:sensitivity}).

The most practical way to improve the $\mu$ results from a {\it PIXIE}-like experiment is to complement it with ground-based observations of the low-frequency synchrotron, free-free, and AME foregrounds. Measuring the synchrotron and free-free SEDs to $0.1\%$ would enable {\it PIXIE} to detect $\mu$ at $\simeq 2\sigma$. 
Some of the challenges could be avoided by selecting specific patches on the sky with low foreground contamination, but we generally find that the distortion sensitivity is mostly limited by the lack of constraints on the {\it shape} of the foreground SEDs rather than their amplitude (Sect.~\ref{sec:foreground_model}). 

Our analysis only uses spectral information to separate different components. Adding spatial information would yield a reduction in the total contribution of fluctuating foreground components (e.g., Galactic contributions) to the sky-averaged spectrum.  In addition, a more optimal sky-weighting scheme could be implemented for the monopole measurement, as opposed to the simple average taken on $70$\% of the sky assumed in our analysis. Extragalactic signals (e.g., CIB) will, however, not be significantly reduced by considering spatial information, unless high-resolution and high-sensitivity measurements become available. 
In this case, extended foreground parameterizations, which explicitly include the effects of spatial averages across the sky and along the line-of-sight \citep{Chluba2017}, should be used. Since {\it PIXIE} has a fairly low angular resolution ($
\Delta \theta\simeq 1.6^\circ$), a combination with future high-resolution CMB imagers might also be beneficial. A more detailed analysis is required to assess the overall trade-offs in these directions.

We close by mentioning that information from the CMB dipole spectrum could also help in extracting the CMB distortion signals \citep{Danese1981, Balashev2015}. In particular, these measurements do not require absolute calibration and thus can also be carried out in the {\it PIXIE} anisotropy observing mode. This can furthermore be used to test for systematic effects. Also, Galactic ($\leftrightarrow$ comoving) and extragalactic foregrounds are affected in a different way by our motion with respect to the CMB rest frame, so that this could provide additional leverage for foreground separation. All this is left to future analysis.

\small
\section*{Acknowledgments}
The authors gratefully thank Al Kogut, Dale Fixsen, and Eric Switzer for their insight and extensive comments, as well as for providing {\it PIXIE} mission parameters and sensitivity. We also thank David Spergel and Jo Dunkley for valuable comments on the manuscript. The authors furthermore cordially thank Daniel Foreman-Mackey for useful discussions about the {\tt emcee} software package. We thank Jacques Delabrouille for noting an error in the {\it COBE/FIRAS} sensitivity curve.
JC is supported by the Royal Society as a Royal Society University Research Fellow at the University of Manchester, UK. This work was partially supported by a Junior Fellow award from the Simons Foundation to JCH.

\begin{appendix}
\section{Parameter Posterior Distributions}

\begin{figure*}
	\includegraphics[width=\textwidth]{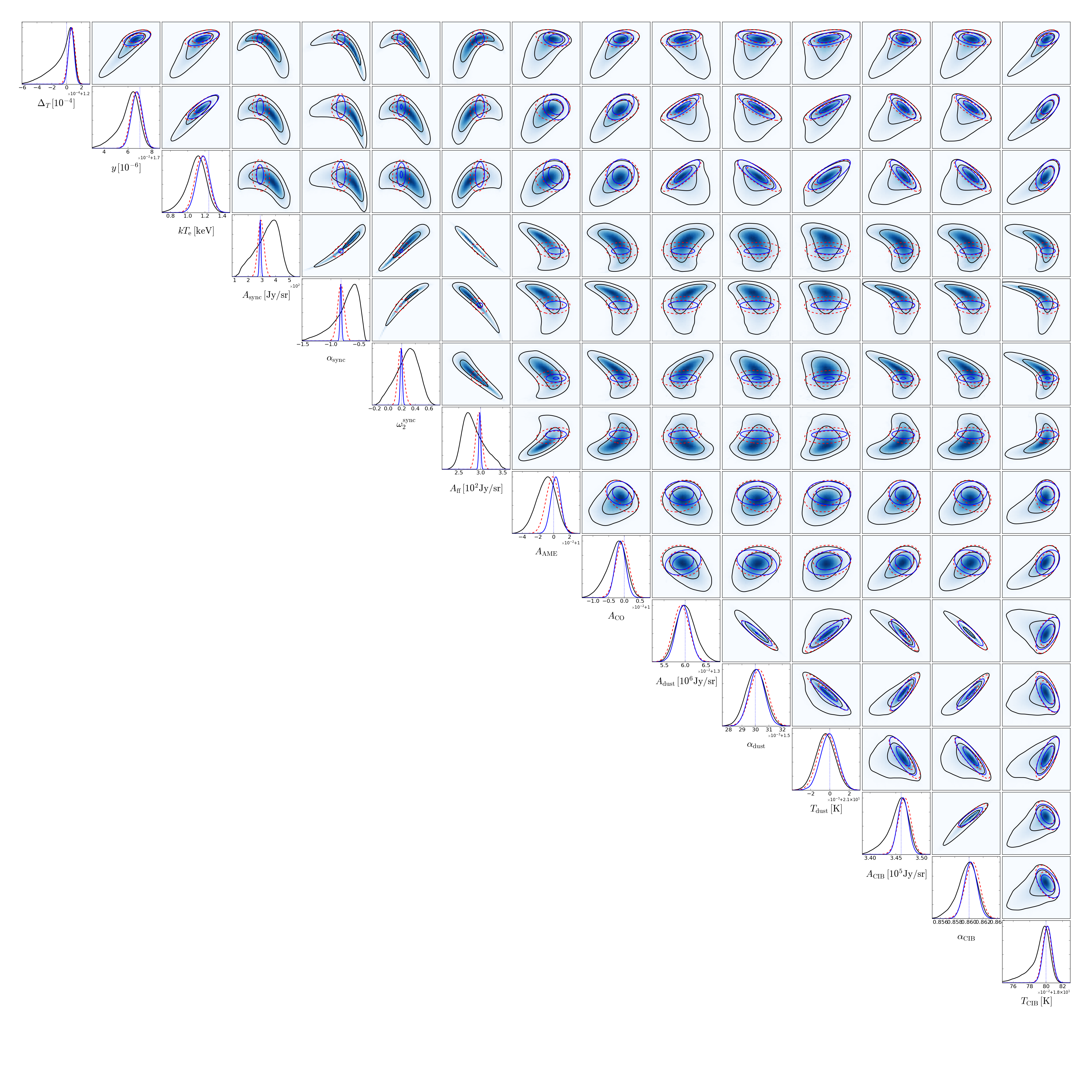}
  \caption{Posteriors obtained with MCMC runs for the full foreground model without $\mu$ in the analysis. The case without priors on the foreground parameters (black line / blue contours) is highly non-Gaussian. Including 10\% priors on $A_{\rm S}$ and $\alpha_{\rm S}$ (red dashed line) leads to much more Gaussian posteriors, improving the CMB parameter constraints by a factor of $\simeq 2$. Tightening the priors on $A_{\rm S}$ and $\alpha_{\rm S}$ to 1\% (blue line) improves the constraints on low-frequency foreground parameters, but only marginally affects the CMB parameter constraints. Details about the error estimates for these cases can be found in Tables~\ref{tab:forecastwpriors} and~\ref{tab:priors}.}
  \label{fig:15p_no_mu_no_prior}
\end{figure*}

\end{appendix}

\small 
\bibliographystyle{mn2e}
\bibliography{Lit}

\end{document}